\documentclass[fleqn,10pt]{wlscirep}
\usepackage[utf8]{inputenc}
\usepackage[T1]{fontenc}
\usepackage{epsfig}
\usepackage{ulem}
\usepackage{graphicx}
\usepackage{caption}
\definecolor{mypurple}{rgb}{0.7,0.3,0.8}
\definecolor{midgreen}{rgb}{0.1,0.6,0.3}
\newcommand{\mjk}[1]{\textcolor{mypurple}{#1}}

\newcommand{\tz}[1]{\textcolor{magenta}{#1}}

\usepackage{cancel}

\author[1]{Jan E. Snellman}
\author[2]{Rafael A. Barrio}
\author[3]{Cecilia I. Ventura}
\author[4]{Tzipe Govezensky}
\author[1,5]{Kimmo K. Kaski}
\author[1,*]{Maarit J. Korpi--Lagg}
\affil[1]{Department of Computer Science, Aalto University School of Science, FI-00076 AALTO, Finland}
\affil[2]{Instituto de F{\'i}sica, Universidad Nacional Aut{\'o}noma de  M{\'e}xico, 01000 M{\'e}xico D.F., Mexico}
\affil[3]{(CONICET) Centro  Atómico Bariloche-CNEA, and Universidad Nacional de Río Negro, 8400-Bariloche, Argentina}
\affil[4]{Instituto de Investigaciones Biomédicas, Universidad Nacional Autónoma de México, México 04510, México.}
\affil[5]{The Alan Turing Institute, 96 Euston Rd, Kings Cross, London NW1 2DB, UK}

\title{Modelling socio-economic effects of the pandemic in a metropolitan area}

\begin{abstract}

One of the most notorious measures taken by authorities in every country  trying to prevent the uncontrollable spread of the Covid 19 pandemic was to impose severe measures to diminish the mobility of people. In recent publications we proposed a mathematical model to address the consequences of these measures and to predict the infections under different scenarios, for various  different countries. In this paper we extend the model, called the BTH--SEIRS-model, to deal with the reduction of mobility in an urban area. The  chosen region was the densely-populated metropolitan area around Helsinki, the capital of Finland.
This choice obeyed to the fact that there are precise data related to mobility from a mobile telephone company, and that this region was completely isolated from the rest of the country for a short period during the pandemic. We also used public transport data from railways and the accommodation occupation rates in the area of interest. We show that the BTH--SEIRS model is capable of simulating the real infection data as well as the mobility-related economic activity, exhibiting the main trends present in the three indicators used for comparison. Our calculations allow us to highlight the difference of the behaviour of the pandemic in a totally different spatial scale, and to discover the hidden factors that govern the behaviour of people who are forced to commute every day from home to work.      

\end{abstract}

\begin{document}

\maketitle

\section{Introduction}
 
It is now known that the Covid-19 pandemic caused almost 7 billion infections and nearly 8 million deaths worldwide \cite{DDG2020}. It has also produced substantial changes in the way we live, conduct our social lives, and care about our physical and mental health and well-being. How people react and behave in such crisis situations depends not only on individual and social factors but also on demographic, cultural, political, societal, and environmental factors of the country they live in. In addition, the reactions of people may change due to receiving real time information about the status of the pandemic, globally and locally. Last, but not least, how the pandemic is spreading in the country and how the country functions during it depends on its socio-economic structure and practices. 

In the early days of the pandemic neither treatments nor vaccines were available, therefore, non-pharmaceutical measures, such as mask wear and mobility restrictions, were implemented to reduce the spread of the infection. The closure of public places, e.g. restaurants, stores, hotels, schools, etc, affected the economy, but not all economic sectors were influenced in the same way. For example, food-related industries were practically not affected, while the pharmaceutical industry was boosted, and many retail areas were adapted using the Internet and messenger services. The most affected were activities related to mobility, such as public transportation, accommodation, tourism, etc. Hence, what happens in a country during a pandemic can be seen as a complex interplay between the dynamics of disease spreading and the dynamics of socio-economic functions in a population-wise and geographically distributed setting.


In order to get insight into this complex system we have developed a mathematical agent-based hybrid model of dynamic disease spreading and socio-economic dynamics of people. It consists of a geographical region divided in cells, where a compartmental SEIRS model (see e.g.\cite{BVGM2013,BKHAG2020,BGBB2020,BVGNBB2021,BGVNBB2022}) dynamically simulates the spread of the disease. Our model also simulates two kinds of agents: authorities implementing mobility restrictions for public health and the population deciding whether they will comply or not with government measures. The SEIRS model is then coupled to the agent-based decision making model, based on the ''better than hypothesis'' (BTH) \tz{ \cite{  SGGBK2017} } of people, to simulate their response to the restrictions set by the authorities. This hybrid model approach has previously been presented \cite{SBKK2022,Somat} and successfully applied to the case of Spain \cite{SpainApp}. 

However, the model has never been applied to a region in which the population density is high and quite uniform, like in a city or urban areas. Therefore, in the study presented here we apply and extend the model to deal with Finland's most densely populated urban area around its capital Helsinki and including the cities of Espoo, Kauniainen and Vantaa, henceforth to be referred to as the "Helsinki metropolitan area" (shown in the map of Appendix \ref{Teliadata}), with a population of about 1.3 million \cite{helsmetropwiki,uusimaaregiongral,statfingral}, which is part of the Uusimaa region of Finland. The Uusimaa region includes 26  municipalities in all\cite{uusimaaregiongral}, and has  a total population of about 1.8 million (representing 31 $\%$ of the  total population of Finland of about 5.65 million), 39 $\%$ of the country's gross domestic product (GDP), and covering 3 $\%$ of the national land area\cite{uusimaaregiongral}. To be precise, the Helsinki metropolitan area, subject of our present work, has  a total population of about 1.3 million (representing 23 $\%$ of the total population of Finland), 32 $\%$ of the country's gross domestic product (GDP), and covers 0,25 $\%$ of the national land area\cite{helsmetropwiki,uusimaaregiongral,statfingral}. This provides a good example of the change of scale of the system we want to study. Furthermore, we had access to the mobile telephone data from the local mobile phone service provider company (Telia), informing the location of each call made by the anonymised local customers in the area, facilitating the present study. In particular, we were able to compare the model´s simulated mobility-related economic activity with real mobility data provided by Telia during the pandemic years. These data are used as a ''virtual experiment'' to provide a comparison with the results from the theoretical model. Earlier studies with epidemic models, using telephone data, include \cite{FMWFM2011,Aleta-et-al-2020,LCWP2021,GDYBGYB2021,TBK2023}. We have also compared the mobility-related economic activity simulated by our model with two indicators obtained from the official Finland database:  StatFin \cite{statfingral}, namely, the quarterly number of passenger railway journeys and the monthly overnight occupation percentage of all accommodation establishments in the Uusimaa region. We found that these two indicators exhibit the same trend, as a function of time during the pandemic, as shown by the Telia data and the simulated mobility-related economic activity.

 Previously, we applied a geo-stocastical SEIRS compartmental model  to study the COVID 19 pandemic  in countries of different sizes, as small as Iceland, medium size as Finland, and large size as Mexico \cite{BKHAG2020} or Argentina \cite{BGBB2020}. In this latter case, we observed that the results were dominated by the metropolitan area of Buenos Aires, which represents a situation of having a large population density in a relatively small geographical region.  
Agent-based epidemic models in city-scale social systems have been based on simulated agents moving between their homes, workplaces, and possibly other establishments (see e.g. \cite{FMWFM2011,SBLAGS2020}). In these models, to mimic a real world situation, the epidemic spreads through the social network of agents. Extending this network approach to a larger and more detailed social systems usually requires very high computing power and special simulation techniques \cite{BBKEXM2008,KCCJHMMYMB2025}, but if one concentrates on a specific aspect of the epidemic, such as its effects on the productivity of firms through supply chains \cite{IMT2021}, it can be more easily accomplished in some cases. The present study provides a test for the limits of our BTH--SEIRS model in small-scale social systems. 

This paper is organised as follows: in section \ref{model} we start by presenting the BTH-SEIRS hybrid model, followed by the data sources used for the mobility-related economic activity and COVID data in Finland (section \ref{datasources}, and Appendix \ref{Teliadata}), ending with the specification of the model parameters and methods used (section \ref{methods}, plus Appendices \ref{biol-parameters} and \ref{simdom}). 
After that we present the main results obtained with our proposed model, and compare them with real Finland data for the pandemic propagation and a series of mobility-related economic activity indicators. We end by the "Discussion" section, and our concluding remarks.

\section{BTH-SEIRS Hybrid Model}
\label{model}

The agent-based BTH-SEIRS model was developed to describe a situation in which an epidemic spreading through a populated geographical region is influenced by the actions of the affected populations to mitigate its spread \cite{SBKK2022}. The mitigation measures by the agents are assumed to directly limit their socio-economic activity, which the agents take into account in their decision making process: the agents choose their level of response by balancing their desire to limit the infection rates and keeping the economic activity levels as normal as possible according to their societal values. Here, the population is divided into two groups of agents (normal population and local authorities) who make decisions about their behavior based on their concerns about health, economic, and social issues. The variables associated with the normal population are in lower case letters, whereas the ones associated with local authorities are in capital letters.

\subsection{BTH method}
\label{BTHpartofthemodel}

The basic dynamical temporal map of the BTH model is
\begin{eqnarray}
x_i &=& \frac{1}{|e_i| + 1} ( \frac{1}{1 - \tilde{w}^c_i} \sum_{k \in e_i} x_k  
 - \frac{1}{1 - \tilde{w}^y_i} (y_i + \sum_{k \in e_i} (y_i - y_k)) 
 + \frac{\tilde{w}^c_i}{1 - \tilde{w}^c_i} ((|e_i| + 1) X^r_i - \sum_{k \in e_i} c_k)),  \label{agentactionx} \\
X_i &=& \frac{1}{|E_i| + 1} (\sum_{k \in E_i} X_k - \tilde{W}^Y_i (Y_i + \sum_{k \in E_i} (Y_i - Y_k)) 
 - \tilde{W}^Z_i (Z_i + \sum_{k \in E_i} (Z_i - Z_k)))
\label{agentactionX}
\end{eqnarray}
where
\begin{equation}
\tilde{w}^y_i = \frac{w^y_i}{|w^x_i|},
\tilde{w}^c_i = \frac{w^c_i}{|w^x_i|}, 
\tilde{W}^Y_i = \frac{W^Y_i}{|W^X_i|}, 
\tilde{W}^Z_i = \frac{W^Z_i}{|W^X_i|}, 
\label{valtilde}
\end{equation}
and $x_j \in [0,1] $ stands for the reduction of economic activity by the agent of normal population $j$,  and $X_k  \in [0,1] $ stands for the restrictions placed by authority agent   $k$. See \cite{SBKK2022,Somat} for details on the restrictions of $x_i$ and $X_i$. The variables $y_j$ and $Y_k$ are the infection rates in the cell of  population agent $j$, and the overall infection rates in the district governed by authority agent $k$, respectively. The quantity  $c_j = x_j - X^r_j$ is the compliance of population  agent $j$ with the measures ordered by the respective authority. In addition, $Z_k = \sum_{j \in D_k } x_j$  is the overall reduction in economic activity in the district $D_k$ governed by the authority agent $k$, respectively. 

Parameters $w^{x,y,c}_i$ and $W^{X,Y,Z}_i$ describe the societal values of the simulated agents. According to the BTH method \cite{SGGBK2017,SBKK2022}, these values are conceptualized as different scales of measure that agents use to draw comparisons between each other. The upper indices refer to the comparison measures, so $w^x_i$, $w^y_i$ and $w^c_i$ measure how much value the population agent $i$ assigns to its own effort to mitigate the epidemic, its own health, and its compliance with the restrictions of the authorities, respectively. Similarly, $W^X_i$, $W^Y_i$ and $W^Z_i$ are the  values assigned by the authority agent $i$ when deciding  the restrictions they put in place to mitigate the epidemic, the overall health and the economic activity in their districts, respectively. It is important to mention here that Eqs. \ref{agentactionx} and \ref{agentactionX} have been derived from the BTH utility functions $U_i$ and $u_i$ introduced for each authority and population agent, respectively, in Ref. \cite{SBKK2022}, as: 
\begin{eqnarray}
U_i &=& \sum_{M} (W_i^{M} (M_i + \sum_{j \in e_i} (M_i - M_j))), \\
u_i &=& \sum_{m} (w_i^{m} (m_i + \sum_{j \in e_i} (m_i - m_j))),
\label{genueq}
\end{eqnarray}
where $M = \{ X,Y,Z \}$ and $m = \{ x,y,c \}$. Under the assumption that all agents make minimum effort to mitigate the spreading of the epidemic (i.e. $U_i=u_i=0$),  parameters $w^x_i$ and $W^X_i$ are required to be negative. Similarly, $w^y_i$ and $W^Y_i$ also need to be negative in order for the agents to want to limit the infection rates in the first place. Similarly, $W^Z_i$ also needs to be negative. Parameters $w^c_i$ do not have similar limitations arising from basic assumptions. We have used this simplification before in \cite{Somat}. 
Therefore, the following three BTH parameter ratios of Eq. \ref{valtilde} are negative: \, $ \, \tilde{w}^y_i < 0 $ , \,  $ \tilde{W}^Y_i < 0 $, \,  $\tilde{W}^Z_i < 0 $. Meanwhile, there are no restrictions for the sign of the compliance ratio: 
$ \tilde{w}^c_i $. We vary the value parameters in time in order to fit the infection  and economic activity data, like in \cite{SpainApp}.

In order for the agents to compare themselves with other agents, in accordance with the BTH spirit, the agents naturally need to be aware of their social neighbourhood. The sets of agents $e_i$ and $E_i$ in Eqs. \ref{agentactionx} and \ref{agentactionX} refer to these neighbourhoods, with the former depicting the set of other population agents that a population agent $i$ can see, while the latter is the set of authority agents seen by authority agent $i$. Due to the limited size of the simulated area, especially in contrast to an entire country like Spain in our earlier study \cite{SpainApp}, we limited the number of authority agents to just one. This means that the restriction equation simplifies to 
\begin{equation}
X = \tilde{W}^Y Y - \tilde{W}^Z Z,
\end{equation}
where $Y$ and $Z$ are the sum totals of the infection rates and economic activity reductionsof all population agents. 

We take $e_i$ to encompass the population agents in a circular area around population agent $i$. The radius of the area is determined by a parameter we denote as the comparison distance ($d_c$), given in units of the lattice parameter of the square grid of population agents modelling the Helsinki metropolitan area. In this work all the population agents have the same comparison distance,  but we had to change it in time during the simulations. We explain these changes in detail in section \ref{methods}.

The model just described has to be coupled with the epidemic model  to simulate the feedback between the actions of the populations reacting to the epidemic and the evolution of the epidemic itself. This allows us to gain insight into the effects that societal values have on the behavior of the population and the spread of the disease. In the BTH--SEIRS model, this coupling is achieved via mobility, as we explain in the following sections.

\subsection{SEIRS compartmental model}
\label{SEIRScom}

 To describe disease spreading over a geographical region we chose the compartmental model approach. The geographical region is divided into square cells of a given size in each of which the disease spreading among its population takes place according to the SEIRS model mechanism between compartments of susceptible (S), exposed (E), infected (I) and recovered (R) \cite{BKHAG2020}. Recovered persons are not susceptible since they attain immunity through infection (natural immunity). To take into account  individuals who attain immunity through vaccination an additional compartment V was added. The dynamics of these compartments  is governed by the following set of equations:  
\begin{eqnarray}
S_{t+1} &=& q (S_t - G_t + \lambda q^b G_{t-b} - \xi S_t) + \mu N \nonumber \\
E_{t+1} &=& q (E_t + G_t - q^{\epsilon} G_{t-\epsilon} ) \nonumber \\
I_{t+1} &=& q (I_t + q^{\epsilon} G_{t-\epsilon}-q^b G_{t-a} ) \nonumber \\
R_{t+1} &=& q (R_t + q^a G_{t-a}-q^b G_{t-b} ) \nonumber \\
V_{t+1} &=& q (V_t + \xi S_t),
\label{SEIRSeq1}
\end{eqnarray}
where
\begin{eqnarray}
q &=& 1 - \mu \nonumber \\
a &=& \epsilon + \omega \nonumber \\ 
b &=& \epsilon + \sigma + \omega \nonumber \\
G_{\alpha} &=& \rho_\alpha S_{\alpha} (1 - e^{-\beta_k I_{\alpha}}),
\label{SEIRSeq2}
\end{eqnarray}
and $\lambda$ is the portion of the population that is susceptible again after having recovered (survival parameter). The parameter $\mu$ is the mortality rate and $\beta_k$ is the transmission parameter of the disease for the variant $k$.  As detailed in Table \ref{table-biol} of Appendix \ref{biol-parameters}, we vary $\beta$  tuning it to the predominant COVID virus variant at each time period in Finland \cite{covariants}.

The latency period $\epsilon$ is the period in which the person is exposed to the virus but is not yet infectious. The period of infectiousness is $\sigma$, and the period of immunity is $\omega$. Regarding the vaccination dynamics $\xi$ describes the effective immunization rate, which is lower than the actual vaccination rate, since experience has shown that not all vaccinated people achieve full immunity \cite{BVGNBB2021}. The probability of being infected within a cell $\alpha$ is calculated as $G_{\alpha}$ where ${\rho_\alpha}$ is the population density in this cell.

The epidemic spreads from infected cells to uninfected cells via the following three dispersion mechanics based on Monte Carlo process \cite{BGBB2020}:
\begin{itemize}

\item An infected cell in direct contact with uninfected cells can infect them one at a time with probability $\nu$, which is a measure of the mobility of people between those cells.

\item Some of the cells are connected via public transport and road networks. The epidemic may spread between any cells connected by these networks with probability $\nu$ weighted by the product of normalized population densities \cite{BGBB2020} of the origin and destination cells (see Appendix \ref{simdom} for details).

\item The epidemic may be spread between distant cells because people perform sporadic  long distance trips. This is considered in our model by a process  similar  to thermal noise. We use a Monte Carlo method\cite{BGBB2020} with a threshold value $ \gamma$ which is explained in Appendix C. 

\end{itemize}
In any of these cases, when a cell is infected  at time $t_0$, the number of infected people is reset by a parameter $\eta$, such that $S_{t_0} = 1 - \eta$ and $I_{t_0} = \eta$.

In addition to these spreading mechanisms, mobility $\nu$ also affects the number of infected people in already infected cells via a mechanism we call "local stochasticity".\cite{BGBB2020} This represents the effect of epidemic mitigation efforts by the authorities and populations successfully reducing the number of infections, and is implemented by setting the number of newly infected in a cell to zero with probability $1-\nu$. Parameter $\nu$ also acts as a bridge between the compartmental and agent based parts of the BTH--SEIRS model, as it is the main variable modified by agent action, as described in the next subsection.

\subsection{Coupling between the two parts of the model}

The coupling between the grid cell populations and the district authorities of the BTH model is accomplished by the dependence of mobility $v_i^t$ on the BTH $x$ variable. For the sake of simplicity, this dependence is assumed to be linear:
\begin{equation}
v_i^t \,  = \,  v_0  \, -  \, x_i \, \,   v_{max},
\label{vtdef}
\end{equation}
where $v_0$ is the maximum  value of the mobility parameter, $0 \leq x_i \leq 1$ is the reduction of the socio-economic activity of the population agent $i$ defined by Eq. \ref{agentactionx}, and $v_{max} < v_0$ is the maximum reduction in the infectiousness of disease that the socio-economic mitigation measures can deliver, since we assume that the population agents will not reduce their economic activity below the level necessary for their survival. See section \ref{methods} and Appendix \ref{simdom} for a discussion of the numerical values of the BTH mobility parameters.

Taking into account that $Z$ represents the pandemic-related reduction of the economic activity of the population agents considered in the  Helsinki metropolitan area  ( $D$ ), in the following we will use as a measure of their total mobility-related economic activity the following quantity:
\begin{equation}
\, A \, \,  =  \, \, 1 \, - \, Z  \, \,  = \,   \, 1 \, - \,  \sum_{j \in D } x_j  \, 
\label{Aeq}
\end{equation}

\begin{figure}[p]
\begin{center}
\includegraphics[width=19cm]{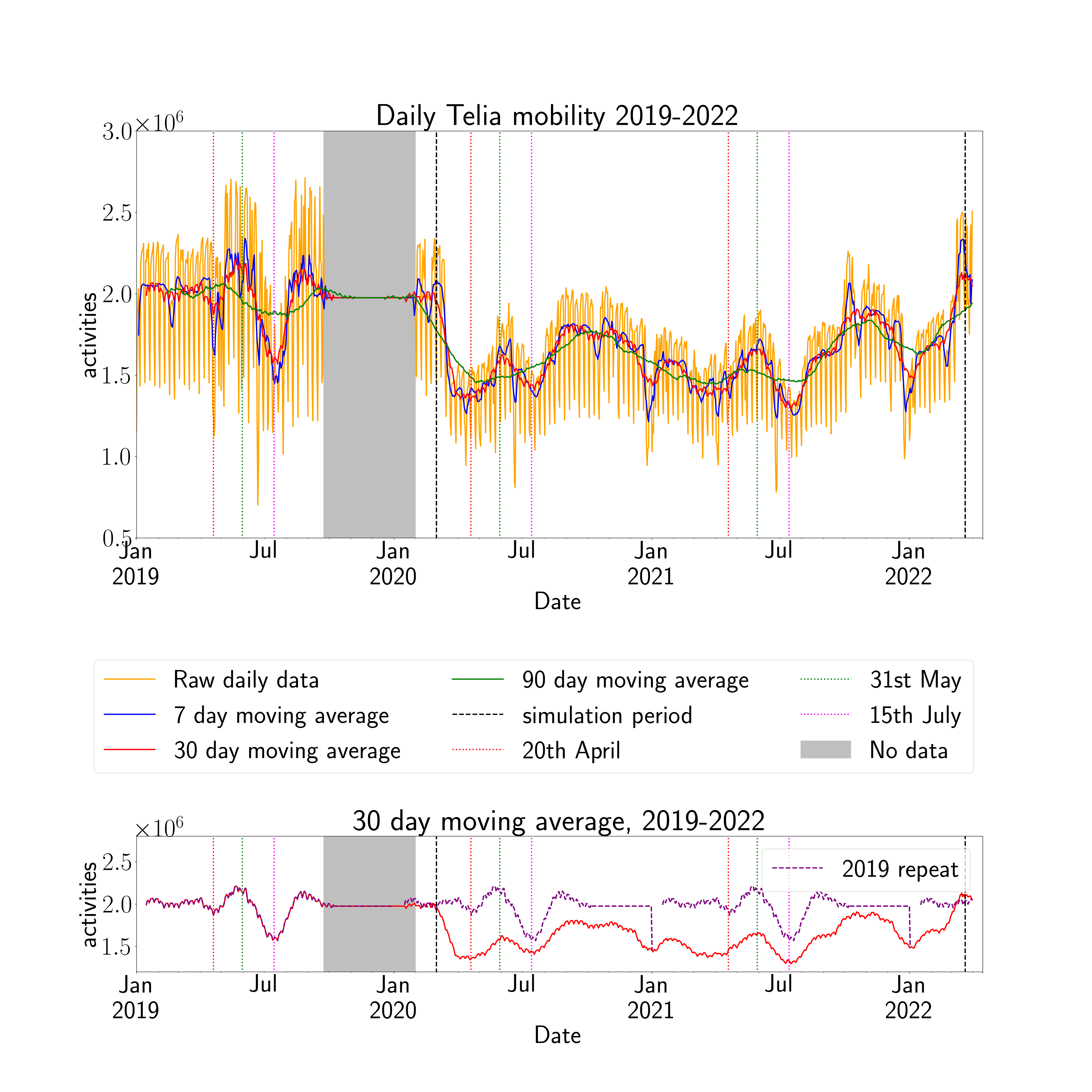}
\end{center}
\caption{The number of total daily activities and its $7$, $30$ and $90$-day moving averages in the Helsinki metropolitan area according to Telia's CrowdInsights database from January 2019 to March 2022 (upper panel), and the comparison of the $30$-day average in the pandemic period  with  the 2019 Telia mobility data (lower panel). The black vertical lines indicate the beginning and the end of the simulation time period, while dotted vertical lines show the seasonal maxima and minima of activities revealed by the $30$-day moving average. The time period between September 2019 and February 2020 was not available in the Telia data: indicated by a gray shaded area. 
} 
\label{OLtel}
\end{figure}

\section{Data sources for mobility-related economic activity}

\label{datasources}

\vskip 10pt

\subsection{Mobility as described by Telia telephone data}
\label{datasources-Telia}

In this section we present the total mobility of people as obtained from the Telia database, purchased from the telephone company Telia by Aalto University (see Appendix \ref{Teliadata}  for the details on the nature of the database), which we later use as an indicator for the mobility-related economic activity, $A$. The data set contains information on where individuals move and  where they are coming from on a daily basis, whereas the BTH--SEIRS model only has a single mobility rate for each geographical cell. This geographical aspect of the Telia data represents a challenge that we had to tackle in the present study, and we chose to use the total mobility as calculated by the total number of visits made by people outside their home territory according to Telia data. 

The total daily mobility according to the Telia database is visualised in Fig. \ref{OLtel}, in which the raw data and some of its moving averages are shown in the upper panel, and the comparison with the 2019 mobility data in the lower panel. The mobility is measured in "activities", which refer to people spending 20 minutes or more away from their designated "home zones", which in turn are designated according to where they have been active before 9:00am in the morning (see Appendix \ref{Teliadata} for details). The dataset does not have any reported data between September 2019 and February 2020, which is indicated by the gray shaded area in the plot. Other than this time period, there are also other days in which data have not been reported, but these are relatively few in number, and we have interpolated over them in the figure.

The usefulness of Fig. \ref{OLtel} is threefold: firstly, it allows us to establish the time frame of the pandemic related disruptions, so that we may define a proper time window for our BTH--SEIRS simulations. Secondly, it helps us to choose the most useful averaging procedure to apply to the Telia-data, as the daily fluctuations are quite large, as clearly seen in the upper panel. And finally, the comparison with the 2019 data in the lower panel helps to visualize eventual  seasonal effects in the average mobility. As for the simulation period, the data show a clear reduction in mobility during the pandemic in 2020 and 2021, with a recovery in 2022. Thus, we decided to focus our BTH--SEIRS simulations on the $750$ day period  after 1 March 2020, which is shortly before the severe pandemic related mobility reduction. In Fig. \ref{OLtel} the chosen simulation period is located between the dashed black vertical lines,  after the gray shaded area (where Telia data are unavailable). In the following figures, we will restrict the time period shown to the indicated simulation period. When choosing the moving averaging time window we considered three options, namely $7$, $30$ and $90$ days, all of which are shown in the upper panel. Considering that the BTH--SEIRS model is only intended to capture the most general of trends in mobility-related economic activity, the most useful averaging time should be long enough to smoothen  daily fluctuations, but not so long that the trends themselves are misrepresented. In the upper panel of Fig. \ref{OLtel} we see that the $7$-day average has still quite strong fluctuations, while the $90$-day average is clearly smoothing out obvious seasonal trends and details of the original daily data. The $30$-day average, however, is free of the influence of strong daily fluctuations, while preserving the relevant trends in the data closely enough. Thus, we chose to use $30$-day moving average in order to smooth the data for the purpose of comparison with the simulation results. 

  \begin{figure}[h!]
  \begin{center}
  \includegraphics[width=14cm]{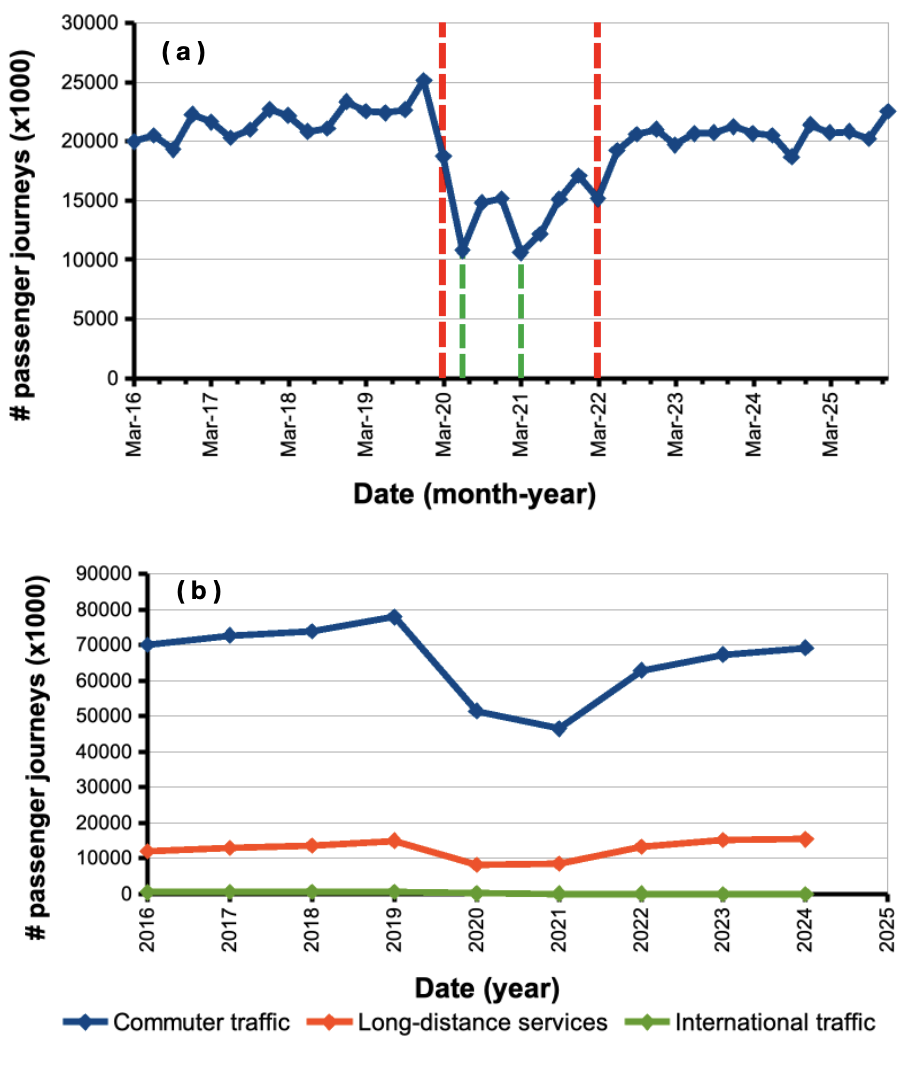}
 \caption{(a)  The total quarterly number of passenger railway journeys in Finland, from year 2016 up to the first semester of 2026. The red vertical lines mark the two-year pandemic simulation period (1 March 2020-2022). Two green vertical lines specify the positions of the main minima: at 1 Jun 2020, and 1 March 2021. 
 (b)  The yearly total number of passenger railway journeys as a function of traffic type in Finland, from year 2016 up to the first semester of 2024. (In that graphic, the country´s long-distance railway data are indistinguishable from the domestic long distance ones.) 
Source: Official Statistics of Finland \cite{statfinR}.}   

    \label{lsfig1}
  \end{center}
  \end{figure}

  \begin{figure}[h!]
  \begin{center}
  \includegraphics[width=14cm]{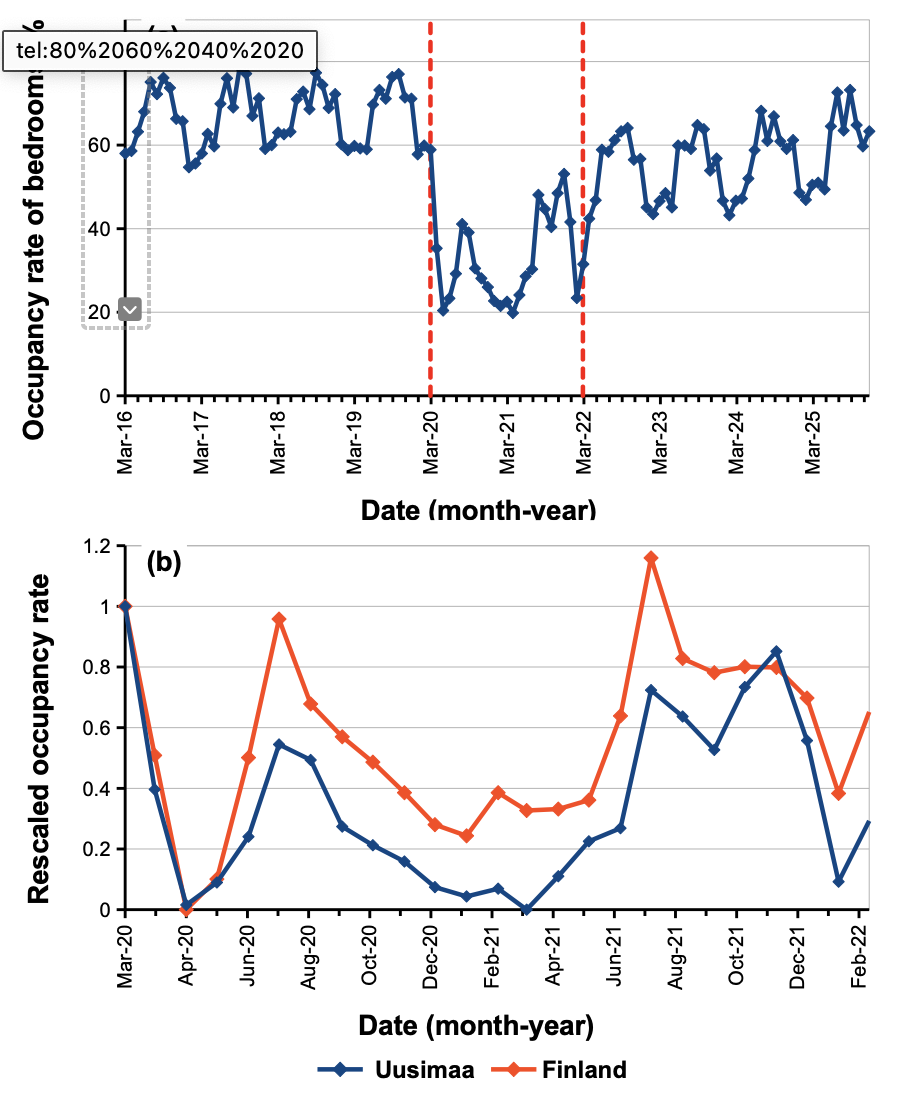}

    \caption[]{(a) The monthly bedroom occupation percentage of all accommodation establishments in the Uusimaa region,  from year 2016  up to the first semester of 2026. The red vertical lines mark the two-year pandemic simulation period (1 March 2020-2022). (b) Comparison of the rescaled  monthly bedroom occupation  data of all accommodation establishments in the Uusimaa region (in blue), with those of the whole country (in red), during the pandemic simulation period. Here, both sets of data were rescaled between the  occupation rate value found at the beginning of the pandemic and the first minima at 1 April 2020.  Source: Official Statistics of Finland \cite{statfinAC}.}

    \label{lsfig2}
  \end{center}
  \end{figure}

\begin{figure}[t]
\centering
\epsfxsize = 0.85\columnwidth \epsffile{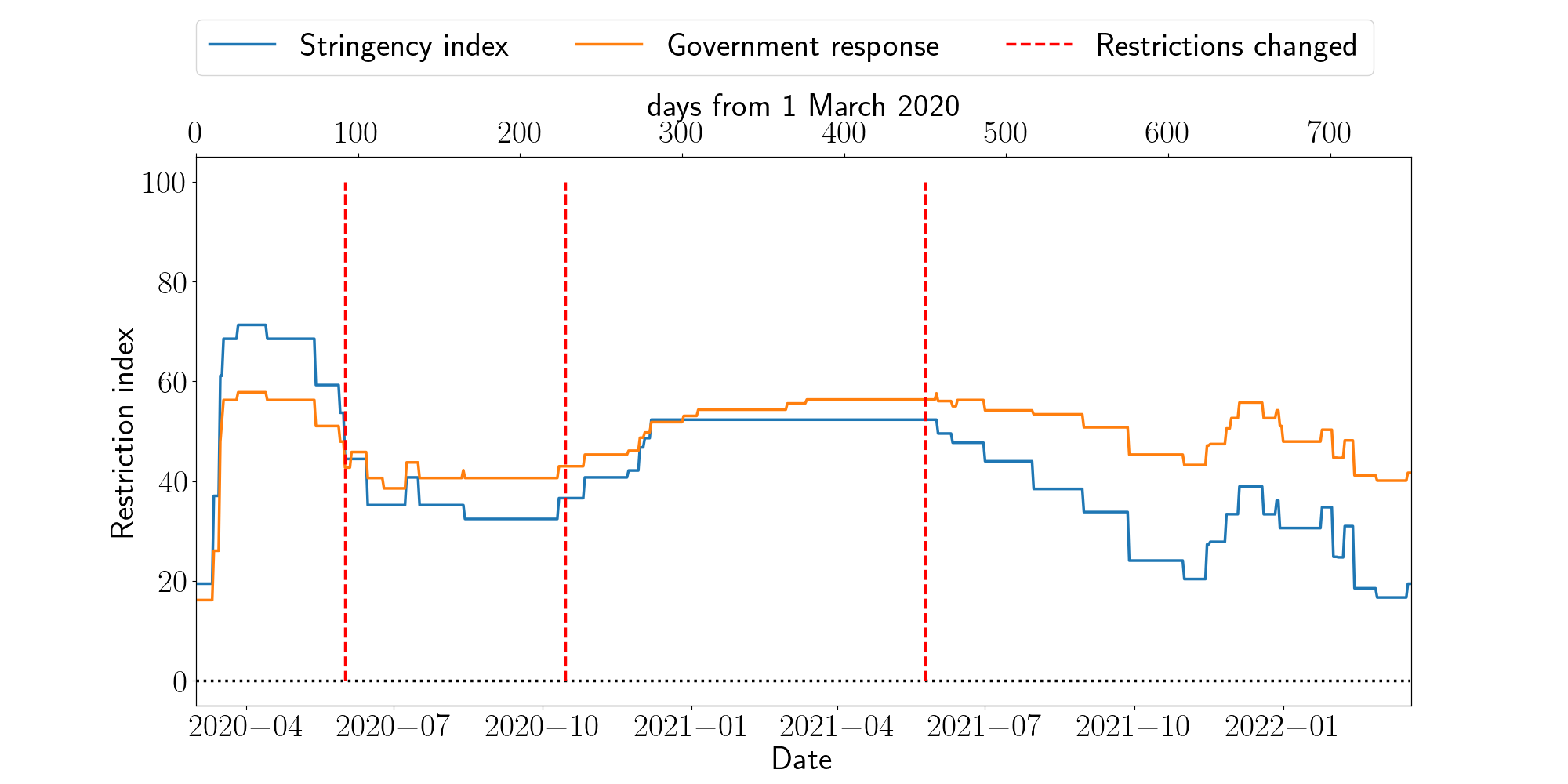}
\caption{Restriction indices for Finland according to Oxford COVID Response Tracker. The time origin (t = 0) corresponds to 1 March 2020. The dashed red vertical lines indicate the three dates (t = 92, 228, 450 days, measured from t=0 i.e.  1 March 2020, or the first of June and tenth of October 2020 and the fifth of May 2021) we chose to change the BTH value parameters of the authority agents in the simulation runs.} 
\label{OLR}
\end{figure}

\vskip 20pt

\subsection{Mobility-related indicators from Official Statistics of Finland database}
\label{datasources-StatFin}

\vskip 20pt

In order to assess  seasonal effects we investigated other indicators from sources  independent of the Telia data, like the  official Finland  StatFin database  \cite{statfingral}. Concretely, in Fig. \ref{lsfig1}(a) we show the total quarterly number of railway passenger journeys before, during and after the COVID pandemic.\cite{statfinR}  The total number of yearly passenger journeys in Finland is dominated by the Uusimaa commuter services, as Fig. \ref{lsfig1}(b) shows.  Notice that  Fig. \ref{lsfig1}(a) exhibits a clear seasonal oscillatory behavior in the quarterly railway data before the pandemic. After that, at the start of the pandemic, a first drastic drop with a minimum at 1 June 2021 is observed, explained by the first set of lockdown measures imposed (as seen in Fig. 5 of Ref.\cite{BKHAG2020}). This is followed by a local maximum between August and October 2020 as restrictions were lifted. Then, as new restrictions were imposed at the end of 2020 due to infections by new variants, by 1 March 2021 a second drop appears, reaching the same minimum value as in June 2020 (as the two green vertical lines in Fig. \ref{lsfig1}(a) stress). After that, as vaccines were available, the number of railway passenger journeys is gradually recovered, slowly tending to levels observed before the pandemic (perhaps affected by other factors such as the Ukraine-Russia war). Notice also, that the clear seasonal oscillatory behavior before the pandemic is not recovered afterwards.

The same main features observed in the railway data are present also in Fig. \ref{lsfig2}, which in panel (a) shows the percentage of bedroom occupation in all accommodation establishments of the Uusimaa region.\cite{statfinAC} These data are monthly, thus providing a higher level of detail than Fig. \ref{lsfig1}(a), but one recognizes very similar trends in both graphs, illustrating the direct relationship between these two indicators obtained from the StatFin database \cite{statfingral} and the mobility. Nevertheless, notice that after the pandemic, though the monthly accommodation data are gradually recovered to a lower level than before the pandemic, here  the clear seasonal oscillatory behavior before the pandemic is recovered afterwards. Panel (b) of Fig. \ref{lsfig2} shows noticeable differences between the monthly  overnight occupation rates in all accommodation establishments of the whole country, and those of the metropolitan Uusimaa region,  during the two pandemic years.\cite{statfinAC} Both  data sets were rescaled so that 1 corresponds to the value at the beginning of the pandemic (i.e. t = 0, or 1 March 2020) and 0 represents the minima found in both sets on April 2020. 
Later we will discuss the origin of these differences in detail, in a wider context.

\vskip 20pt
\subsection{COVID data sources}

In addition to the data on the railway travel, accommodation usage, and telephone activity, we also made use of the restriction indices reported by the Oxford COVID Response Tracker \cite{Thomas:2021} shown in Fig. \ref{OLR} to define the government mobility restriction changes. The stringency and government response indices plotted in the figure are aggregates of the total of $56$ different pandemic related variables, which provide concise information on the overall public and official measures and responses to COVID. For future reference, the figure shows the dates,  t = 92, 228, 450 days, measured from t=0 i.e.  1 March 2020,  where the BTH government value parameters are changed (three dashed vertical lines  at:  1 June and 10 October 2020, and 5 May 2021). Open data from the Finnish Institute for Health and Welfare provided us with the vaccination rates to use in our simulations \cite{THLvac}, as well as numbers of COVID infections confirmed by testing \cite{THLinf} to compare our simulation results with.

\section{Parameters and Methods}
\label{methods}

Our hybrid model enables us to separate different aspects of the problem. Namely, for the SEIRS part of the model (see Sect.~\ref{SEIRScom}), we could use the same biological parameters of our previous work \cite{BKHAG2020,BVGNBB2021,BGVNBB2022},  taking into account the specific COVID variants present in Finland during the pandemic (for details, see Appendix \ref{biol-parameters}). In our  simulations, we divided the whole Helsinki metropolitan land area in 1km by 1km square cells. Next, we will focus on the BTH parameter values used (defined in Sect.~\ref{BTHpartofthemodel}). In our  simulations, for the whole Helsinki metropolitan land area we take  only one authority agent, and one population  agent per 1km by 1km cell, thereby including a total of 877 population agents. Further details on how the simulation domain and the transport network mentioned in Sect.~\ref{SEIRScom} are defined, can be found in Appendix \ref{simdom}.

\begin{figure}[ht!]
\centering
\epsfxsize = 1\columnwidth 
\epsffile{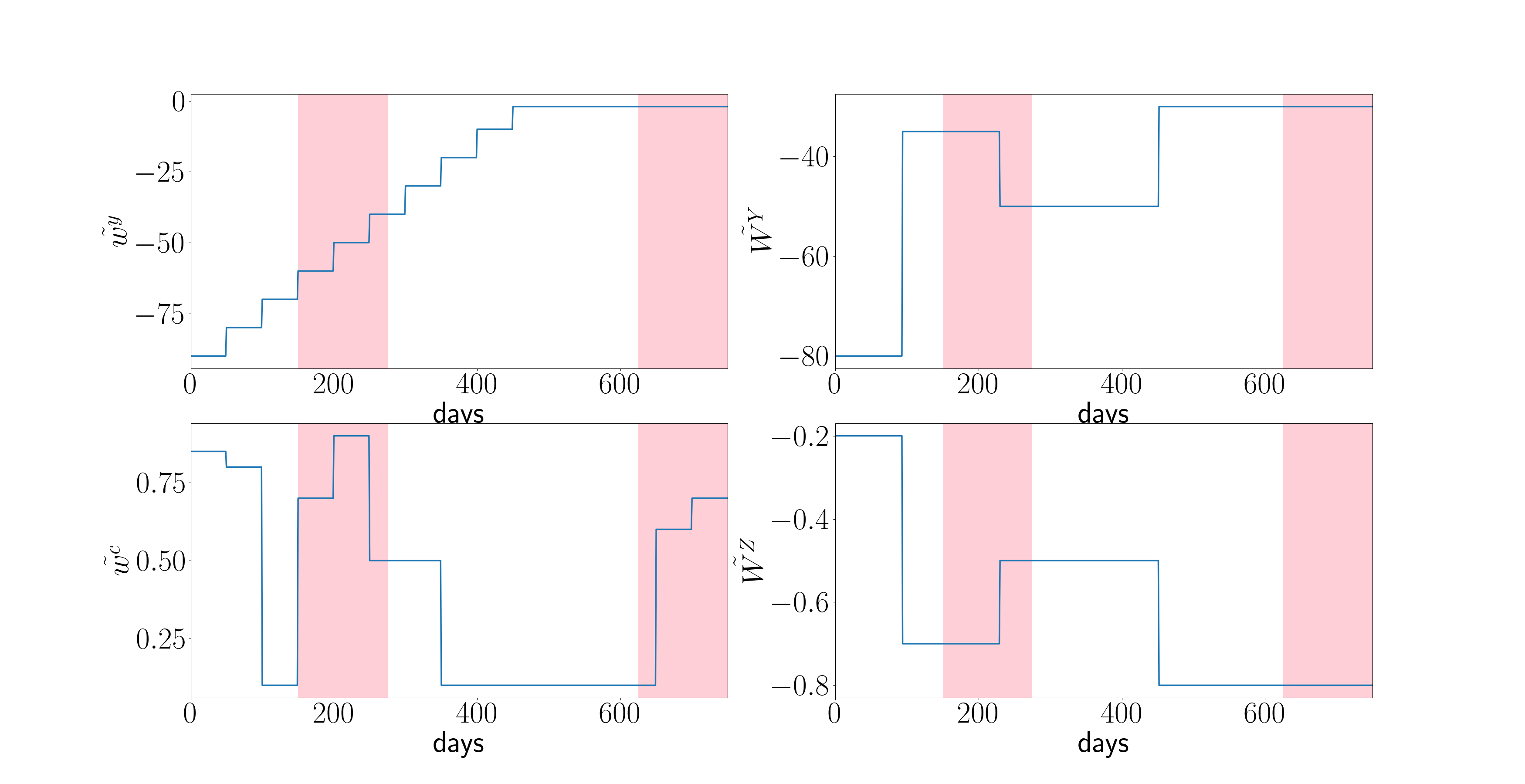}
\caption{Optimal ratios of BTH-value parameters (as defined in Eq.~\ref{valtilde}) as a function of time, providing the best fit to the simulated infection cases and economic activity. Panels on the right: reflect the mobility restriction changes by the authorities at t= 92, 228, 450 days (shown in Fig.~\ref{OLR}). The background colour refers to the time epochs, when different values of comparison distance are used in our simulations: white corresponds to $d_c \approx 10$km (global awareness period) , while in the shaded pink areas: $d_c \approx 4$km (local awareness period ). The specific dates corresponding to the $d_c$ changes are: $ t = 150, 275, 625$ days. t = 0 corresponds to 1 March 2020. }
\label{OLval}
\end{figure}

As mentioned in Sect.~\ref{BTHpartofthemodel} the following agents' weights: $w_x$, $w_y$, $W_x$, $W_y$ , and $W_z$ were considered negative to meet the minimum effort assumptions. As shown in Eq.~\ref{valtilde}, the BTH--equations depend only on  four relevant  ratios of the original six BTH-value parameters, which are restricted to having negative signs, except for the compliance ratio $\tilde{w}^c_i$ as discussed in Sect.~\ref{BTHpartofthemodel}. It was assumed that  the compliance ratio $\tilde{w}^c_i$   should be positive, considering that the general population was initially willing to comply with the restrictions imposed by the authority agents. In Fig.~\ref{OLval} we show the four BTH parameter ratios used in the simulations presented in next section, assuming they are equal for all population agents. 

\begin{figure}[h] 
\centering
\epsfxsize = 1.2\columnwidth \epsffile{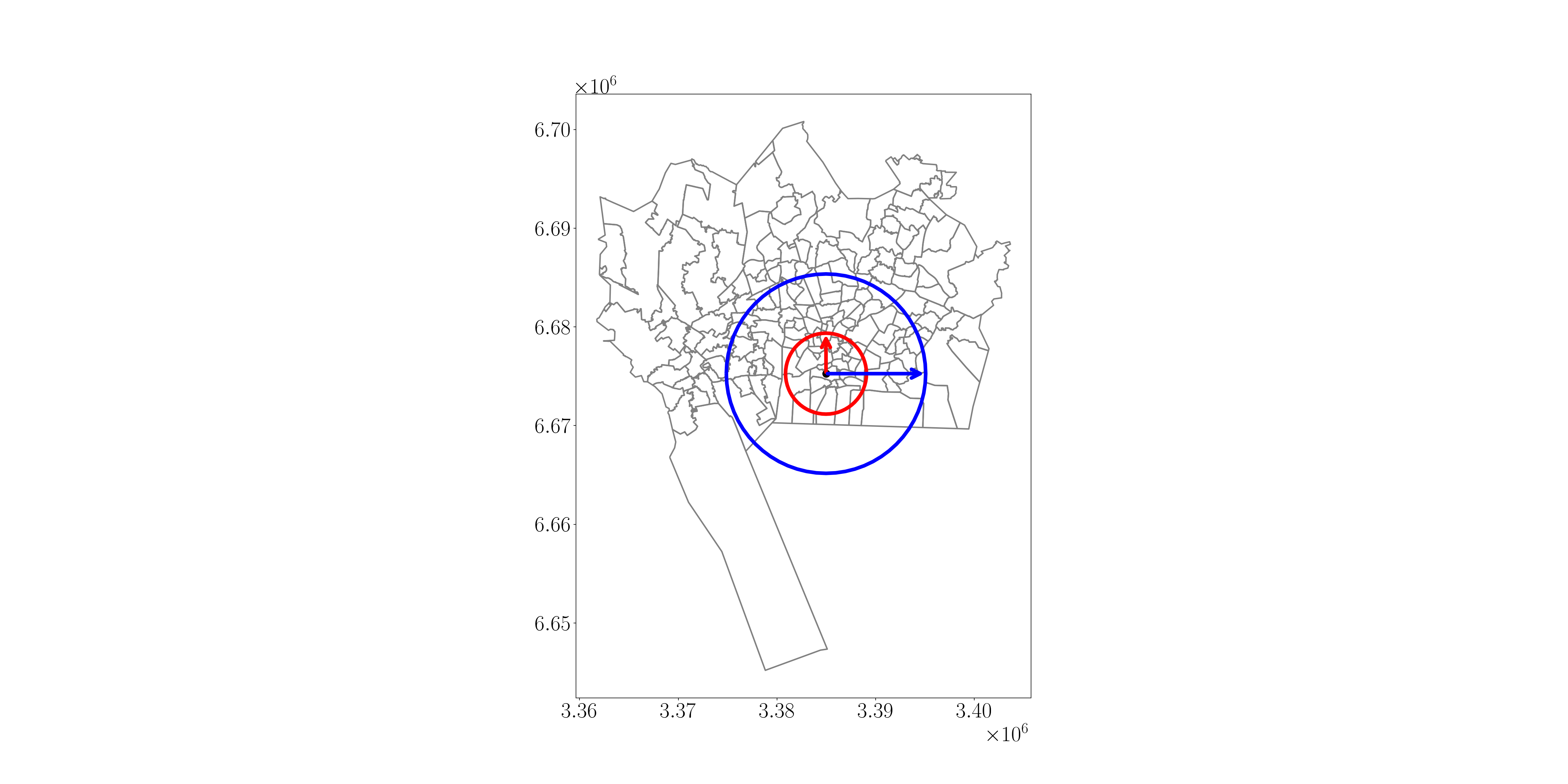}
\caption{Visualization of the two comparison distances $d_c$, used in Fig.~\ref{OLval}, compared with the whole simulated region. The  larger (blue) radius is $d_c \approx 10$km (used during the global awareness period), while the smaller (red) circle radius equals $d_c \approx 4$km (used during the local awareness period).} 
\label{CDvis}
\end{figure}

Notice  that we have assumed changes in the values of the population agent parameters to occur every $50$ days, and let the health value parameter $w^y_i$  (shown in top left panel) steadily tend to its highest value of $-2.0$ from its initial value of $-90.0$. This trend represents the change of COVID from an unknown new disease to something more familiar as time goes on. Meanwhile, we have based the parameters of the government's health and economic values $\tilde{W}^Y_i$ and $\tilde{W}^Z_i$ on the restriction indices reported by the Oxford COVID Response Tracker shown in Fig.~\ref{OLR}. The scale of $\tilde{W}^Y_i$ is kept similar to $\tilde{w}^y_i$, while the scale of $\tilde{W}^Z_i$ is kept closer to the scale of $\tilde{w}^c_i$, which is kept between the values $0$ and $1$.

The days in which we chose to change the BTH value parameters of the authority agents in the simulations  are shown as dashed red vertical lines in Fig.~\ref{OLR}, indicating these three dates: t = 92, 228, 450 days \cite{Thomas:2021}, measured from t=0 i.e.  1 March 2020. Notice that during the first three periods up to day 450, the restrictions reported by the Response Tracker are relatively stable, though in the fourth and final period there are some fluctuations, probably related to the wave of Omicron variant of the virus. To simplify, we have chosen to represent the authority agent value parameters in each period as constants. The r.h.s. panels of Fig.~\ref{OLval} exhibit opposite behaviours, because we assume that in periods of higher health concerns, the authorities will prioritize them over economic activity concerns, and vice versa. Having fixed most of the BTH value parameters as discussed, leaves  only the compliance ratio $\tilde{w}^c_i$ as a free parameter. We have used it to fit the mobility-related economic activity and infection rates, as discussed in next section.  

The comparison distance $d_c$  for each population agent $i$, introduced in Sect~\ref{BTHpartofthemodel}, was defined as the grid size of the social neighbourhood this agent uses to decide its eventual reduction of economic activity ($x_i$) and compliance to the regulations imposed by the authorities ($c_i$), in presence of the pandemic. Due to the smaller scale of the Helsinki metropolitan area studied in this work, in our present simulations we found that we needed to assume  two values of the comparison distance $d_c$,active at different time periods, shown by respective  different background colours in Fig.~\ref{OLval}, either white or pink.  We call these two states of comparison distance: local and global awareness. Concretely, here the Helsinki area population agents have two different states that reflect how many agents in their vicinity they are aware of when making decisions. We found that comparison distances needed to be chosen so that they reach $10$ and $4$ neighbours in cardinal directions ($7$ and $2$ in the diagonal directions) in order to to  fit our simulations to the real epidemiological and economic data presented in next section. In the scale of our grid, these correspond to approximately $10$ and $4$ kms in real distances. Namely, up to day t=150, $d_c \approx 10$ kms, which means that each population agent is aware of a large proportion of  the Helsinki metropolitan area  population agents, as shown in Fig.~\ref{CDvis} (by the blue circle). This global awareness appears at the breakout of the pandemic, when due to lockdown restrictions the population was informed by TV, radio, internet, etc.,  of global data of the evolution of the pandemic. Later, between t = 150 to 275 days, as seen in Fig. \ref{OLval}, comes a period characterized by a smaller comparison distance, $d_c \approx 4$ kms, reflecting the fact that during this epoch, the population agents' decisions ceased to be strongly dependent on the global lockdown measures by the authorities, which were lifted around the end of this time period. After that, the awareness and actions of the population agents turned more towards the concerns in their local population. Later, new virus variants present from t=275 to t=625, lead to a renewed larger scale of awareness of the population, consistent with  a period of more concern about higher infection rates. After t=625, the summer of 2021 and the vaccination completed lead to a new period of closer interactions between population agents, and smaller scale awareness.

\vskip 20pt

\section{Results}
\label{newresul}

\begin{figure}[p]
\begin{center}
\includegraphics[width=15cm]{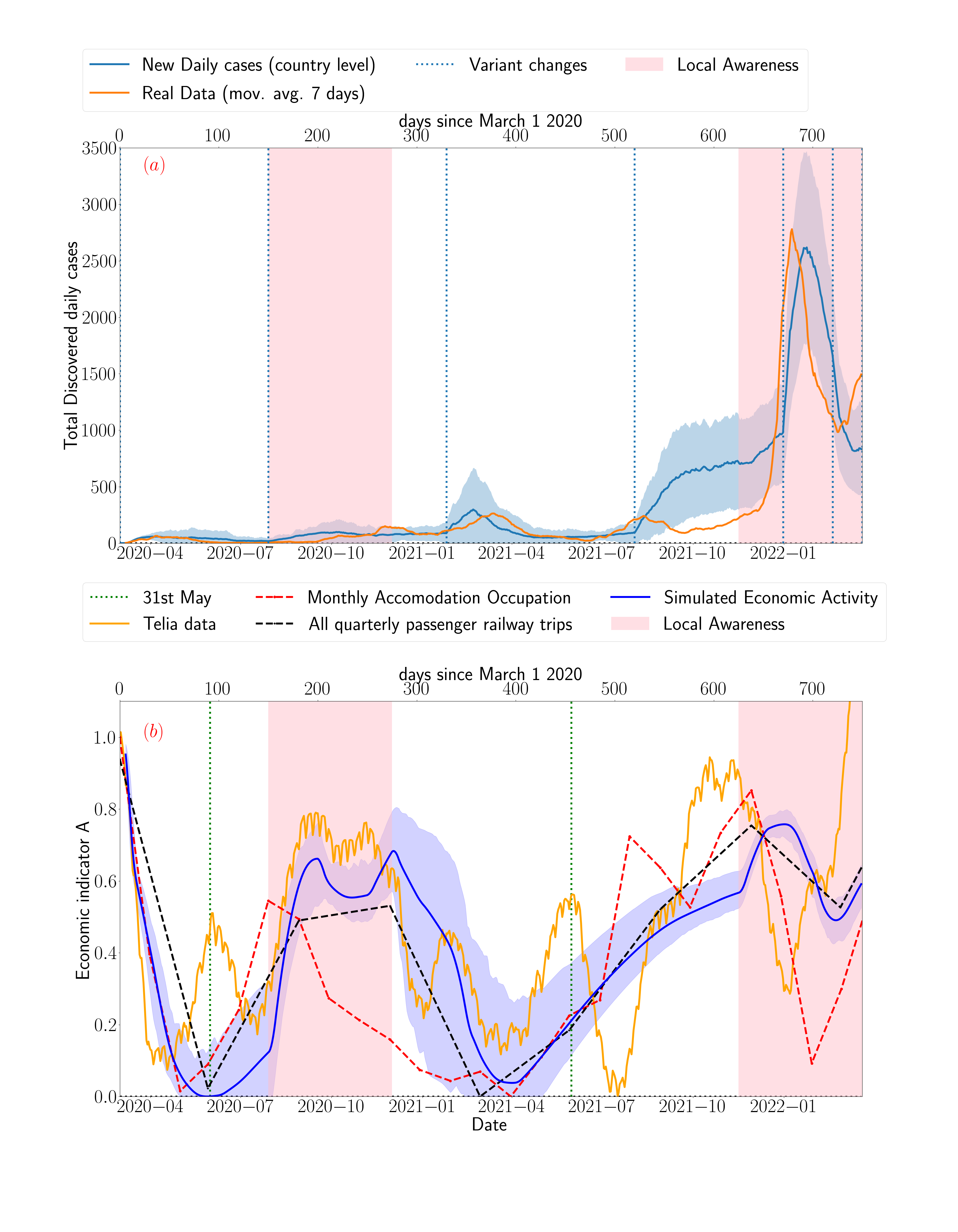}
\end{center}
\caption{ (a) BTH--SEIRS simulated infection results compared to the confirmed infection cases \cite{THLinf}. (b) BTH--SEIRS simulated mobility-related economic activity, $A$ (Eq.~\ref{Aeq}), with its estimated error (shaded in lilac), compared to these three Helsinki metropolitan area indicators: the daily Telia telephone data, the quarterly number of railway journeys \cite{statfinR}, and the monthly accommodation occupation \cite{statfinAC}. The background colour refers to the two values of comparison distance used in our simulations: white corresponds to $d_c \approx 10$km, while in shaded pink areas: $d_c \approx 4$km. Seasonal maxima in the Telia data, discussed in the text,  are indicated as dotted green vertical lines. 
$t = 0$ corresponds to 1 March 2020. }
\label{OLInf}
\end{figure}

In this section we present the results obtained with our proposed BTH-SEIRS model for the mobility-related economic activity in Finland during the pandemic years, and the corresponding COVID infection cases. Concretely, Fig.~\ref{OLInf} focuses on the pandemic years. We present the results  for the COVID daily infection cases in the upper panel, and for the mobility-related economic activity in the lower panel.Real data are also included for comparison: in Fig.~\ref{OLInf}(a) we show the real 30-day averaged Telia mobility data and in Fig.~\ref{OLInf}(b) we additionally include the monthly accommodation occupation data and the quarterly railway data, presented in Sect.~ \ref{datasources}. We rescaled all the data shown in Fig.~\ref{OLInf}(b) to the range between $0$ and $1$,  so that the value $1$ corresponds to each economic indicator at $t=0$, i.e. 1 March 2020 here. Our results for the economic activity and infection rates  are based on the average of $100$ realizations and the error estimates are based on removing 5\% of the highest and lowest realizations, respectively, so that the shaded areas in Fig.~\ref{OLInf} contain 90\% of the realizations. 

In Fig.~\ref{OLInf} one recognizes that effectively the same trends are present in the results of our simulations and the superposed real data included for comparison.  The results demonstrate that with the BTH--SEIRS model it is possible to simulate the mobility-related economic activity and describe the Telia, railway and accommodation data quite accurately. Notice that at the same time, we are able to fit the infection rates during the first half of the simulation period, 
while sacrificing some accuracy in later waves when other variants are present.

Notice that during the pandemic there are also differences between the three mobility-related indicators exhibited in Fig.~\ref{OLInf}(b), related to the specific nature of each one. Namely, the monthly Uusimaa all-establishment occupation percentage reflects the accommodation used by non-Uusimaa residents, and therefore exhibits clear seasonal effects modulated by the presence of the disease. For example, there are clear maximal accommodation  values during  both  summer seasons, while in winter:  a pronounced  decrease of the Uusimaa accommodation occupation is observed from  fall-winter 2020 until reaching, near $t = 400$  (corresponding to March 2021, as seen in Fig.~\ref{lsfig2}(b)), a minimum with  almost zero rescaled accommodation occupation (corresponding to a 20 $\%$ occupation), just as in the initial phase of the pandemic, at $t=60$. As evident in Fig.~\ref{lsfig2}(b), this is not the case in the data for the whole of Finland \cite{statfinAC}: where the first accommodation occupation minimum  corresponds to  13 $\%$ at April 2020 (t = 60 days), while the minimum reached one year later is noticeably higher, namely 23 $\%$ at January 2021 (t = 340 days). This winter 2020-2021  minimum is due to the appearance of new disease variants, without yet available vaccination. After that winter, both the Uusimaa region and the whole of Finland exhibit a clear recovery of their accommodation data, but the effect is faster and more  pronounced for the whole country, reaching maxima in summer by July 2021 as Covid vaccination already reached the majority of the population of both regions.

Meanwhile, Fig.~\ref{OLInf}(b) shows that the main difference between the monthly  accommodation data (mainly used by non-Uusimaa residents)  and the  quarterly total number of railway journeys where the Uusimaa commuter traffic prevails (see Fig.~\ref{lsfig1} of Sect.~\ref{datasources-StatFin}) appears during fall-winter 2020. This is due to the fact that commuter trains are mainly used by Uusimaa residents for work and daily activities (observing non-pharmaceutical measures after the lockdown restrictions). At the beginning of the pandemic, there is also a noticeable difference in the sharp decrease in railway data that seemingly appeared with a delay due to the decrease in monthly accommodation data, which is only due to the quarterly nature of the StatFin railway data.

Focusing now on the 30-day averaged Telia data included for comparison in Fig.~\ref{OLInf}(b), the Telia-described mobility decrease during  the first quarter of 2020  only differs from the corresponding monthly-accommodation and quarterly-railway decreases, due to the fact that the Telia original data are daily. The same origin explains the sequence of minima observed in the three indicators after day t=650 in Fig.~\ref{OLInf}(b): namely the minima observed in daily Telia data precedes the one of monthly accommodation data, with a larger delay present in the quarterly railway data. Seasonal effects are well described by the daily Telia data, as one can see in the comparison with year 2019 in the lower panel of Fig. ~\ref{OLtel}. This explains the Telia local maxima near days 100 and day 450, and the local minimum at day 500,  not apparent  in the  accommodation and railway data of Fig.~\ref{OLInf}(b). 

Taking into account the origin of the differences between the three indicators discussed above, it is interesting to stress that the mobility-related economic activity  simulated by our BTH-SEIRS model fits most accurately the railway data after day t=100 ( the difference in the previous days is due to the delay already discussed due to the quarterly nature of the StatFin railway data used). We attribute this connection to the fact that the mobility described the Uusima railway public transport, dominated by Uusimaa residents, is affected by the disease propagation in a similar way to the BTH-SEIRS mobility description. Consistent with this interpretation, being the accommodation data mostly related to non-Uusimaa residents, there is a more indirect  connection which explains the difference with the mobility-related economic activity described by our model. Finally, the 30-day averaged daily Telia data are more detailed, but they are also including mobility data related to non-public transport: where the impact of the disease is not so direct. This further explains why the  economic activity data of Fig.~\ref{OLInf}(b) obtained with the BTH-SEIRS model, based directly on the disease propagation, are better described by the public transport railway data. 
Notice also that the Telia data include specific seasonal details, which our BTH-SEIRS model is not directly taking into account. 

In contrast to our present application of the BTH-SEIRS model to the  homogeneously and densely populated Helsinki metropolitan area of Finland,   it is interesting to point out that 
the first time  we applied the  model  to a realistic situation, 
we had focused on a whole country, Spain\cite{SpainApp}, with a very inhomogeneous geographical distribution of its population \cite{SpainApp}. In that case, we found that a single constant value of comparison distance during the pandemic  sufficed to fit the only economic indicator which was used to compare the simulated main mobility-related economic activity of Spain. Namely, the accommodation occupation percentage impacted by the government lock-down measures. Our choice was also justified by the fact that the tourism sector represents about 10-12 $\%$ of the GDP product of Spain, with about 
55 $\%$ of the total tourism expenditure linked to international visitors.   

Concretely, to describe  Spain's infection evolution and economic activity related to tourism  a square lattice grid of BTH-population agents, with lattice parameter of $5 kms$,  and a number of regional authorities were used.\cite{SpainApp} In that simulation, the mobility-related economic activity could be fitted during the pandemic  with a  comparison distance of  $d_c = 7.5 kms $, meaning that each population agent $i$ takes into account the behaviour of its eight nearest neighbours \cite{SpainApp}. Nevertheless, in common with the present application to the Helsinki metropolitan area, we did have to vary the BTH-value parameters for the Spanish authority and population agents over time, in order to reproduce the infection  and tourist accommodation data. We conjecture that a single comparison distance for all population agents was found to be sufficient for Spain, since the economic activity simulation was performed using a nation-wide monthly accommodation indicator, which includes an averaging process of the regional data. Focusing on the same indicator for Finland, Fig.~\ref{lsfig2}(b) clearly exhibits the difference with the data of a single metropolitan region of the country (Uusimaa), which is due to the averaging effect involved for the whole country data.

In fact, when averaging properties of an inhomogeneous system, like a whole country, the dispersion around the mean value is large, blurring local details. Instead, a smaller dispersion is obtained when averaging a homogeneous system, like a metropolitan area,  thus allowing to describe local details. If we add time dependent interactions between the system components,  a large dispersion might also blur their effects (which might be the case for a whole country). Meanwhile, a smaller dispersion might preserve a time dependent modulation of the interactions between the system components. This might be the case for population agents of metropolitan densely-populated areas, where the government mobility restrictions or health concerns by new COVID variants would provide the time-modulation of the interactions between agents, leading to different levels of awareness of their neighbours (i.e. the need to diversify the comparison distance, the simplest scenario being to assume two values, like in our simulations for the Helsinki metropolitan area). Also, in an urban area where the population density is higher and more uniform, the local stochasticity mechanism becomes dominant, greatly reducing the responsiveness of the model to changes in value parameters. Using two different comparison distances turns out to be simplest way to describe this phenomenon.

\section{Discussion}

Here we have presented a BTH-SEIRS model to describe the evolution of the COVID pandemic and its impact on mobility-related economic activity, adapted to the densely populated Helsinki metropolitan region of Finland, which includes the capital of Helsinki and represents 32$\%$ of the country's GDP. The model takes into account not only the properties and characteristics of the disease, but also the social aspects of the attitude of people and authorities facing a pandemic. The model allows us to examine the interplay between the social behaviour of people  and disease in different environments and cultural reactions (e.g. compliance to mobility-restrictive measures by the authorities). It also provides tools to examine the system on very different scales, such as worldwide, a geographical region, single countries, and even a single urban region, as is the case of the present study.

The BTH-SEIRS model used is a hybrid model including agent-based and epidemiological features, in addition to a geographical description of the region, including the population density distribution. The SEIRS compartmental part of the model describes the evolution of the pandemic. Two kinds of agents, respectively representing the regional authorities and the population, react to the pandemic. The authorities impose restriction measures on the mobility of the population agents, to contain the propagation of the pandemic, thereby also  affecting the economic activity.  Meanwhile the population agents decide their actions according to their main concerns. They change their mobility taking into account the epidemiological situation, deciding their compliance to the eventual authorities' restriction measures in place, considering also the response of other population agents and the economic situation.  The general BTH--SEIRS model was presented in 2022 \cite{SBKK2022,Somat}, and its first realistic application was to the country of Spain \cite{SpainApp} in 2024. 

Here we adapted the BTH-SEIRS model to treat a new realistic situation: the description of the situation of the Helsinki metropolitan region during the two COVID pandemic years, regarding the epidemic evolution and the mobility-related economic activity, in the context of government mitigation measures and the social response of the population. For the SEIRS part of the model, the COVID virus biological parameters were taken from established medical data as before\cite{BKHAG2020,BVGNBB2021}, considering also the main variants in Finland during the pandemic\cite{covariants} (as detailed in Appendix \ref{biol-parameters}) . It was challenging to tune the BTH-SEIRS model to a region of such a small scale: concretely, with a land area 657 times smaller than that of Spain, but at the same time with a population much more homogeneously and densely distributed geographically, e.g. the mean population density in the Helsinki metropolitan area is about 17 times that of Spain. As a first step, we decided to represent the Helsinki metropolitan area government by a single BTH-authority agent. Meanwhile, a total of 877 BTH-population agents in a square lattice grid of 1 km x 1 km cells  are distributed to cover the Helsinki metropolitan land area. 

To face the challenge of tuning the model to the scale represented by the Helsinki metropolitan area, here we used three different mobility-related economic activity indicators (or proxies) during the two pandemic years. The more detailed indicator used for comparison with our present model simulations consisted of the 30-day average of the daily mobile telephone data obtained from the Uusimaa service provider, Telia, reflecting directly the mobility of the local customers through the precise location of their movements. The other two mobility-related indicators used for comparison  were obtained from the official StatFin national database\cite{statfingral}. One of them was the monthly Uusimaa accommodation occupation percentage of all lodging establishments ( similar to the only indicator used for comparison when applying the model to Spain). The third and final indicator used here for mobility was the quarterly total national number of railway journeys which is dominated by the Uusimaa commuter traffic. 

It is interesting to stress that  differences between the three chosen mobility indicators due to their inherent nature, allowed us to assess to what extent  the BTH-SEIRS model is able to describe different aspects of the system under study.

After tuning the BTH-SEIRS parameters for the model simulations of the Helsinki area pandemic evolution and the mobility-related economic activity during the pandemic, one recognizes that effectively the same trends of the real data described by the chosen indicators are present. Though one might conclude that the model is correctly simulating the Helsinki area mobility-related economic activity in general, it is evident that the best described indicator is the quarterly number of railway journeys. This is due to the more direct connection between public transport used by the Uusimaa population for mobility-related economic activities and the spread of the pandemic, which is the precise connection assumed between the BTH and the SEIRS parts of the model. In contrast, Telia mobility data take into account all means of transportation:  e.g. also people walking alone, or using private cars, which affect much less the pandemic propagation. Meanwhile, due to the smaller scale of the Uusimaa region, the accommodation indicator describes the bedroom occupation of lodgings mainly by non-Uusimaa residents, therefore with a less direct connection to the mobility-related economic activity and disease status of the Uusimaa population.  

It is interesting to focus on the effect of the difference of scale exhibited by the accommodation occupation indicator: while in a country like Spain, it provided a very good basis to tune the BTH-SEIRS model, in a metropolitan area like the Helsinki region it is less relevant to produce a good overall fit. Furthermore, if one would aim to fit the BTH-SEIRS model for a smaller scale region, like the Helsinki metropolitan area, using  a single mobility-related indicator:  the best results would be  achieved using a monthly public transport indicator. A further advantage of having used three indicators for the mobility-related economic activity simulation in the Helsinki metropolitan area, is that we could compare the relevance of seasonal effects. As discussed in detail in the Results section, these are prominent for the Telia and accommodation data, while less relevant in public transport. Not surprisingly, our BTH-SEIRS model exhibits less seasonal effects. This is due to the fact that,  in its present form, the proposed model does not include seasonal effects directly. 

Furthermore, another effect of the small scale problem  addressed was the need to diversify the size of the group of neighbour population agents which one of them takes into account to decide e.g. its compliance to authority measures, or disposition to reduce its economic activity, etc. Concretely, two comparison distances were found to be necessary to describe, at different stages of the pandemic,  periods characterized by either global awareness (when serious pandemic concerns are present. As information of the pandemic spread reaches the population on a global scale,  population agents  reduce their mobility: either due to  lock-downs being in place,  or by self-preservation, when new unknown pandemic waves strike and vaccination is not yet readily available). Meanwhile,  local awareness to their surrounding population agents, characterizes periods in which the population relaxes their medical concerns. 

A straightforward application of our tuning of the BTH-SEIRS model for the Helsinki metropolitan region, would be to predict the  evolution of an epidemic or pandemic caused by a different virus, and its effect on the mobility-related economic activity of this region, just by changing the biological parameters of the SEIRS part of the model. Also,  the use of the three chosen indicators for the Helsinki metropolitan area, as well as tuning the model,  enabled us to gain a deeper understanding of the capabilities  and limitations of the BTH-SEIRS model in its present form to describe details of a smaller scale region like  metropolitan Helsinki.  This provides a good starting point for future applications of the model by adapting it to different scenarios. For example,  applying it to a country with some predominant metropolitan regions having populations which differ in their main mobility-related economic activities, or other characteristics ( e.g. population density distribution, education possibilities, main means of transportation, cultural differences, etc), which might require to differentiate the BTH parameters of the metropolitan areas.  

\vskip 20pt

\section{Conclusions}

The model presented here takes into account not only the properties and characteristics of the disease but also the social aspect of the attitude of people and authorities facing a pandemic. It is important to note that the model could be adapted to deal with different spatial scales and urban distributions, as it was shown in its application to the case of a large country (Spain) and a relatively small-size metropolitan area (Helsinki). Furthermore, the model allows one to examine the interplay between the social behaviour of people and disease in different environments and cultural reactions (compliance with restrictive measures).

As a matter of fact, the small scale of the metropolitan area subject of this study  required a diversification of the comparison distance, i.e. the size of the neighbour group taken into account by each population agent to decide its social behaviour and mobility. Depending on the 
perception of people of the health risks versus their economic and social needs, periods with global awareness ( characterized by a larger comparison distance, coincident with periods where health concerns dominate), were followed by periods with local awareness (where smaller groups of neighbours are taken into account by population agents, now less preoccupied by health issues and able to resume their usual economic and social activities).  

 Finally,  having superimposed the previously shown railway and accommodation data with the Telia mobility data during the pandemic years, one recognizes that the same trends are effectively present. Therefore, one can also see that our model is correctly simulating mobility-related economic activity. We were able to adjust the mobility parameters of the model fitting all data, which allows us to discover crucial aspects of the social behavior of a population with very distinct cultural habits facing potentially a quite catastrophic phenomenon of a pandemic. 

\section*{Acknowledgements}

We thank Dr Nadia Barreiro for creating the population density and road network maps used in the simulations and for many helpful conversations regarding the topics of this study. The computational resources provided by the Aalto Science-IT project have been utilised in this work. KK acknowledge support from EU HORIZON 2020 INFRAIA-1-2014-2015 program project (SoBigData: Social Mining and Big Data Ecosystem) No. 654024 and INFRAIA-2019-1 (SoBigData++:European Integrated Infrastructure for Social Mining and Big Data Analytics) No. 871042 as well as NordForsk Programme for Interdisciplinary Research project "The Network Dynamics of Ethnic Integration". R.A.B., J.E.S. and N.B. acknowledge support from The National Autonomous University of Mexico (UNAM) and Alianza UCMX of the University of California (UC), through the project included in the Special Call for Binational Collaborative Projects addressing COVID-19.  M.J.K. acknowledges funding from the European Research Council (ERC) under the European Union's Horizon 2020 research and innovation programme, through project ``UniSDyn'', grant agreement n:o 818665.

\bibliography{BTHgeneral}

@Article{Somat,
author="Snellman, Jan E.
and Barrio, Rafael A.
and Kaski, Kimmo K.
and Korpi-Lagg, Maarit J.",
year="2024",
date="2024/12/01",
title="A modelling study to explore the effects of regional socio-economics on the spreading of epidemics",
journal="Journal of Computational Social Science",
page="2535-2562",
volume="7",
issue="3",
issn="2432-2725",
url="https://doi.org/10.1007/s42001-024-00322-2",
doi="10.1007/s42001-024-00322-2"}

@Article{SpainApp,
author="Snellman, Jan E.
and Barreiro, Nadia L.
and Barrio, Rafael A.
and Ventura, Cecilia I.
and Govezensky, Tzipe
and Kaski, Kimmo K.
and Korpi-Lagg, Maarit J.",
year="2024",
date="2024/01/08",
title="Socio-economic pandemic modelling: case of Spain",
journal="Scientific Reports",
page="817",
volume="14",
issue="1",
issn="2045-2322",
url="https://doi.org/10.1038/s41598-023-44637-y",
doi="10.1038/s41598-023-44637-y"}

@INPROCEEDINGS{FMWFM2011,
  author={Frias-Martinez, Enrique and Williamson, Graham and Frias-Martinez, Vanessa},
  booktitle={2011 \- IEEE \- Third \- International \- Conference \- on \- Privacy, \- Security, \- Risk \- and \- Trust \- and \- 2011 \- IEEE \- Third \- International \- Conference \- on \- Social \- Computing}, 
  title={An Agent-Based Model of Epidemic Spread Using Human Mobility and Social Network Information}, 
  year={2011},
  volume={},
  number={},
  pages={57-64},
  doi={10.1109/PASSAT/SocialCom.2011.142}}

@article{DDG2020,
title="An interactive web-based dashboard to track {COVID-19} in real time",
author="Dong, Ensheng
and Du, Hongru
and Gardner, Lauren",
year="2020",
doi="doi: 10.1016/S1473-3099(20)30120-1",
journal="The Lancet Infectious Diseases",
pages="533-534",
volume="20",
issue="5",
publisher="Elsevier"
}

@article{SBKK2022,
title = {Modelling the interplay between epidemics and regional socio-economics},
journal = {Physica A: Statistical Mechanics and its Applications},
volume = {604},
pages = {127696},
year = {2022},
issn = {0378-4371},
doi = {https://doi.org/10.1016/j.physa.2022.127696},
url = {https://www.sciencedirect.com/science/article/pii/S0378437122004629},
author = {Jan E. Snellman and Rafael A. Barrio and Kimmo K. Kaski and Maarit J. Käpylä}
}

@article{BVGNBB2021,
      title={Strategies for {COVID-19} vaccination under a shortage scenario: a geo-stochastic modelling approach}, 
      author={N. L. Barreiro and C. I. Ventura and T. Govezensky and M. Núñez and P. G. Bolcatto and R. A. Barrio},
      year={2022},
      journal = {Scientific Reports},
      Volume = {12},
      Pages = {1603},
      DOI = {https://doi.org/ 10.1038/s41598-022-05481-8}
}

@TechReport{IMT2021,
  author={INOUE Hiroyasu and MURASE Yohsuke and TODO Yasuyuki},
  title={{Do Economic Effects of the Anti-{COVID-19} Lockdowns in Different Regions Interact through Supply Chains?}},
  year=2021,
  month=Jan,
  institution={Research Institute of Economy, Trade and Industry (RIETI)},
  type={Discussion papers},
  url={https://ideas.repec.org/p/eti/dpaper/21001.html},
  number={21001},
  doi={},
}

@article{BKHAG2020,
      title={Modelling {Covid-19 epidemic in Mexico, Finland and Iceland}}, 
      author={Rafael A. Barrio and Kimmo K. Kaski and Gudmundur G. Haraldsson and Thor Aspelund and Tzipe Govezensky},
      year={2021},
      journal = { Physica A},
      volume = {582},
      pages = {126274}
}

@article{BGBB2020,
      title={Detecting infected asymptomatic cases in a stochastic model for spread of {Covid-19. The case of Argentina}}, 
      author={Nadia L. Barreiro and Tzype Govezensky and Pablo G. Bolcatto and Rafael A. Barrio},
      year={2021},
      journal = {Scientific Reports}, 
      volume = {11}, 
      pages = {10024}, 
      DOI = { https://doi.org/10.1038/s41598-021-89517-5 }
}

@article{SBLAGS2020,
title = "COVID-ABS: An agent-based model of COVID-19 epidemic to simulate health and economic effects of social distancing interventions",
journal = "Chaos, Solitons \& Fractals",
volume = "139",
pages = "110088",
year = "2020",
issn = "0960-0779",
doi = "https://doi.org/10.1016/j.chaos.2020.110088",
url = "http://www.sciencedirect.com/science/article/pii/S0960077920304859",
author = "Petrônio C.L. Silva and Paulo V.C. Batista and Hélder S. Lima and Marcos A. Alves and Frederico G. Guimarães and Rodrigo C.P. Silva"
}

@article{BVGM2013,
author = {Barrio, Rafael and Varea, Carmen and Govezensky, Tzipe and José, Marco},
year = {2013},
month = {01},
pages = {2143-2176},
title = {Modeling the Geographical Spread of Influenza {A(H1N1): The Case of Mexico}},
volume = {7},
journal = {Applied Mathematical Sciences},
doi = {10.12988/ams.2013.13193}
}

@article{SGGBK2017,
author = {Snellman, Jan E and I{\~n}iguez, Gerardo and Govezensky, Tzipe and Barrio, R A and Kaski, Kimmo K},
title = {Modelling community formation driven by the status of individual in a society},
journal = {Journal of Complex Networks},
volume = {5},
number = {6},
pages = {817-838},
year = {2017},
doi = {10.1093/comnet/cnx009},
URL = { + http://dx.doi.org/10.1093/comnet/cnx009},
eprint = {/oup/backfile/content_public/journal/comnet/5/6/10.1093_comnet_cnx009/5/cnx009.pdf}
}

@article{Thomas:2021,
   title = {A global panel database of pandemic policies (Oxford {COVID-19} Government Response Tracker)},
   author = {Hale, Thomas and Angrist , Noam and Goldszmidt, Rafael and Kira, Beatriz  and Petherick, Anna  and Phillips, Toby and Webster, Samuel and Cameron-Blake , Emily and Hallas, Laura and Majumdar, Saptarshi and Tatlow, Helen},
   journal = {Nature Human Behaviour.},
   volume = {5},
   pages = {529–538},
   year = {2021},
   DOI = {https://doi.org/10.1038/s41562-021-01079-8}

}

@article{BGVNBB2022,
    author = { Barreiro, N. L. and Govezensky, T and Ventura, C. I. and Núñez, M. and Bolcatto, P. G. and  Barrio, R. A. },
    title = {Modelling the interplay of {SARS-CoV-2} variants in the {United Kingdom}},
    journal = {Scientific Reports},
    pages ={12372},
    year = {2022},
    DOI = {https://doi.org/10.1038/s41598-022-16147-w}
}

@Misc{covariants,
    author = {Emma B. Hodcroft},
    title = {CoVariants: {SARS-CoV-2} Mutations and Variants of Interest},
    year = {2021},
    note = {\url{https://covariants.org/}}
}

@Misc{statfingral,
    author = "{StatFin}",
    title = "{ Official Statistics of Finland (OSF) database [online publication].}",
    year =  "{2026}",
    note = {https://stat.fi/en/statistics/matk. url: https://pxdata.stat.fi/PxWeb/pxweb/en/StatFin/. }
}

@Misc{uusimaaregiongral,
    author = "{Helsinki-Uusimaa Regional Council-Official Statistics 
    database. [Online publication]}",
    title = "{}",
    year =  "{2026}",
    note = {https://uudenmaanliitto.fi/en/helsinki-uusimaa-region/ }
}

@Misc{helsmetropwiki,
    author = "{Helsinki-Metropolitan area data, Wikipedia 5 June 2026. [Online publication]}",
    title = "{}",
    year =  "{2026}",
    note = {https://en.wikipedia.org/wiki/Helsinki-metropolitan-area/
     }
}

@Misc{statfinAC,
    author = "{Helsinki: Statistics Finland}",
    title = "{Official Statistics of Finland (OSF): Accommodation statistics [online publication] Access method:
https://pxdata.stat.fi/PxWeb/pxweb/en/StatFin/StatFin-matk/11ix.px/chart/chartViewLine/.}",
    year = "{[Referenced: 17.7.2026]}",
    note = {ISSN=1799-6325. }
}

@Misc{statfinR,
    author = "{Helsinki: Statistics Finland}",
    title = "{Official Statistics of Finland (OSF): Railway statistics [online publication]
    Access method: 
https://pxdata.stat.fi/PxWeb/pxweb/en/StatFin/StatFin-rtie/12ra.px/chart/chartViewLine/.}",
year = "{[Referenced: 17.7.2026]}",  
    note = {ISSN=2670-3351. }
    }

@Misc{THLinf,
author = "{Finnish Institute for Health and Welfare (2020-2026) [Electronic Database]}",
title = "{Confirmed corona cases in Finland (COVID-19)}",
year = "{[Referenced: 2.2.2024]}",
note = {https://sampo.thl.fi/pivot/prod/api/epirapo/covid19case.json}
}

@Misc{THLvac,
author = "{Finnish Institute for Health and Welfare(2020-2026) [Electronic Database]}",
title = "{Progress of COVID-19 vaccinations}",
year = "{[Referenced: 15.3.2024]}",
note = {https://sampo.thl.fi/pivot/prod/api/vaccreg/cov19cov.json}
}

@ARTICLE{Aleta-et-al-2020,
author = {Aleta, A. and Martín-Corral, D. and Pastore y Piontti, A. and Ajelli, M. and Litvinova, M. and Chinazzi, M. and Dean, N. E. and Halloran, M. E. and Longini, Jr I. M. and Merler, S. and Pentland, A. and Vespignani, A. and Moro, E. and Moreno, Y.},
year = {2020},
date = {2020/09/01},
title = {Modelling the impact of testing, contact tracing and household quarantine on second waves of COVID-19},
journal = {Nature Human Behaviour},
pages = {964-971},
volume = {4},
issue = {9},
ISSN = {2397-3374},
url = {https://doi.org/10.1038/s41562-020-0931-9},
doi = {10.1038/s41562-020-0931-9}}

@ARTICLE{LCWP2021,
author = {Levin, R. and Chao, D. L. and Wenger, E. A. Proctor, J. L.},
year = {2021},
date = {2021/09/01},
title = {Insights into population behavior during the COVID-19 pandemic from cell phone mobility data and manifold learning},
journal = {Nature Computational Science},
pages = {588-597},
volume = {1},
issue = {9},
ISSN = {2662-8457},
url = {https://doi.org/10.1038/s43588-021-00125-9},
doi = {10.1038/s43588-021-00125-9}}

@article {worldpop:fin,
author={Bondarenko, M. AND Priyatikanto, R. AND Tejedor-Garavito, N. AND Zhang, W. AND McKeen, T. AND Cunningham, A. AND Woods, T. AND Hilton, J. AND Cihan, D. AND Nosatiuk, B. AND Brinkhoff, T.AND Tatem, A. AND Sorichetta, A.},
title={2025 Constrained estimates of 2015-2030 total number of people per grid square at a resolution of 30 arc (approximately 1km at the equator) R2025A version v1. Global Demographic Data Project},
year={2025},
journal={WorldPop},
note={Funded by The Bill and Melinda Gates Foundation (INV-045237). WorldPop - School of Geography and Environmental Science, University of Southampton. DOI:10.5258/SOTON/WP00840}
}

@article{GDYBGYB2021,
    doi = {10.1371/journal.pone.0253865},
    author = {Guan, Grace AND Dery, Yotam AND Yechezkel, Matan AND Ben-Gal, Irad AND Yamin, Dan AND Brandeau, Margaret L.},
    journal = {PLOS ONE},
    publisher = {Public Library of Science},
    title = {Early detection of COVID-19 outbreaks using human mobility data},
    year = {2021},
    month = {07},
    volume = {16},
    url = {https://doi.org/10.1371/journal.pone.0253865},
    pages = {1-17},
    number = {7},

}

@INPROCEEDINGS{BBKEXM2008,
  author={Barrett, Christopher L. and Bisset, Keith R. and Eubank, Stephen G. and Xizhou Feng and Marathe, Madhav V.},
  booktitle={SC '08: Proceedings of the 2008 ACM/IEEE Conference on Supercomputing}, 
  title={EpiSimdemics: An efficient algorithm for simulating the spread of infectious disease over large realistic social networks}, 
  year={2008},
  volume={},
  number={},
  pages={1-12},
  doi={10.1109/SC.2008.5214892}}

@INPROCEEDINGS{KCCJHMMYMB2025,
  author={Kitson, Joy and Costello, Ian and Chen, Jiangzhuo and Jiménez, Diego and Hoops, Stefan and Mortveit, Henning and Meneses, Esteban and Yeom, Jae-Seung and Marathe, Madhav V. and Bhatele, Abhinav},
  booktitle={2025 IEEE International Parallel and Distributed Processing Symposium (IPDPS)}, 
  title={Pandemics in Silico: Scaling Agent-Based Simulations on Realistic Social Contact Networks}, 
  year={2025},
  volume={},
  number={},
  pages={484-496},
  doi={10.1109/IPDPS64566.2025.00050}}

@ARTICLE{TBK2023,
  
AUTHOR={Takko, Tuomas  and Bhattacharya, Kunal  and Kaski, Kimmo },
         
TITLE={Modelling exposure between populations using networks of mobility during COVID-19},
        
JOURNAL={Frontiers in Physics},
        
VOLUME={Volume 11 - 2023},

YEAR={2023},

URL={https://www.frontiersin.org/journals/physics/articles/10.3389/fphy.2023.1138323},

DOI={10.3389/fphy.2023.1138323},

ISSN={2296-424X}}

\newpage

\appendix

\section{Telia data}
\label{Teliadata}

In this work, we use the Telia database purchased from the Telia telephone company by Aalto University, which concerns the movement of people within a small portion of southern Finland in the vicinity of the capital city of Helsinki. Namely the Helsinki metropolitan area shown in  Fig. \ref{TelData}. The figure allows to  visualize the data collection method: 

\begin{figure}[h]
\centering
\epsfxsize = 0.85\columnwidth \epsffile{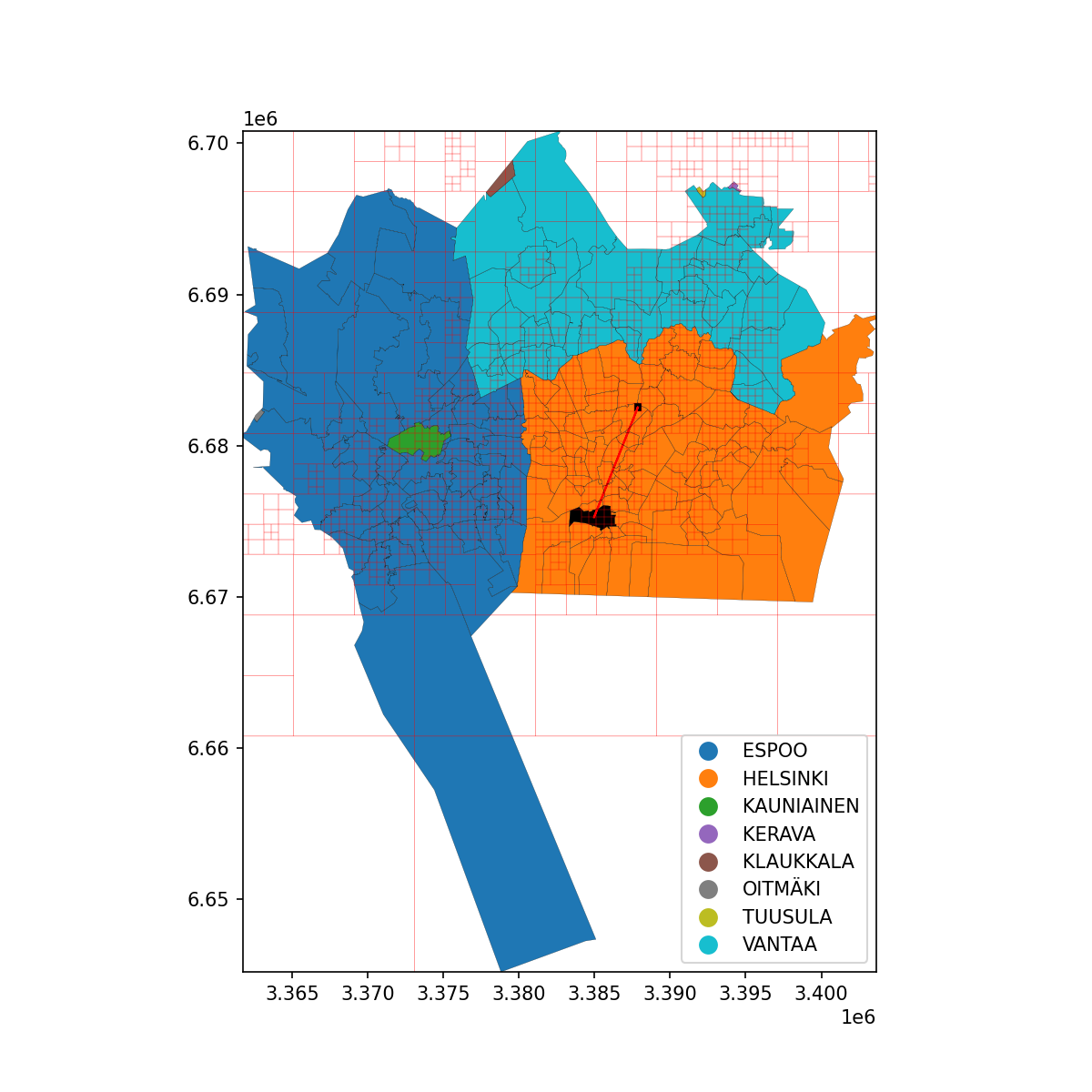}
\caption{A visualisation of the Telia data. The postcode areas are separated by black boundaries, while the Telia grid cells are separated by red boundaries. The coloured areas represent cities or municipalities as indicated. }
\label{TelData}
\end{figure}

\begin{itemize}

\item On each day, people with Telia cell phone plans are given designated "home zones" according to their activities before 9:00 in the morning, or first their pings. The location of this zone is recorded at the accuracy level of postcode areas, at most.

\item When people move during the day, their location is tracked by the Telia telephone grid, which consists of squares of varying sizes according to local population numbers. If a person with a certain designated home zone stays in a given Telia grid cell for more than 20 minutes, this stay is recorded as an activity within that cell.

\item For privacy reasons, if there are fewer than $5$ people recorded as coming from a certain home zone to a given Telia grid cell, that specific data  is not reported in the Telia dataset. 

\item Days with low signal quality are not reported in the data set.
    
\end{itemize}

As an example, the postcode area $"00100"$ is highlighted as black in Fig. \ref{TelData} and connected by a red line to a randomly chosen Telia grid cell, which is also coloured black. The number of people who start their days in this postcode area and later in the day spend more than $20$ minutes in the highlighted cell is recorded as the number of activities for the day for this combination of postcode area and the Telia cell. 

\newpage

\section{SEIRS epidemiological parameters in Finland }
\label{biol-parameters}

\vskip 10pt

The first panel of Table \ref{table-biol} specifies all biological SEIRS parameters of the model. It includes the immunization parameter $ \xi$ and the disease transmissitivity $\beta$, whose time dependence has been adapted according to Finland data as detailed in the two lower panels. Virus-specific parameters like the latency period ($\epsilon$),  the infectious period ($\sigma$), the immunity period ($\omega$), the survival parameter ($\lambda$) and the threshold to start an infectious focus ($\eta$) are independent of the country.

In the lower panel of Table \ref{table-biol} we specify the different $(k)$ virus variants which were more relevant in Finland, and their effect on the transmission parameter $\beta$.  The dates specified for the change of $\beta$ values are chosen to coincide approximately with the appearance of a new dominant variant in Finland ( representing at least 50 percent share of the total infections), as reported by CoVariants  \cite{covariants}. Of the $\beta$ values shown in Table \ref{table-biol} the first five entries are based on our earlier works \cite{SpainApp}, only the last one is new to this particular study. This innovation was done in order to model better the aftermath of the omicron wave in Finland , even though we do not simulate far beyond the large omicron peak. Also, $\eta$ parameter was adapted to the higher population density by multiplying it by ten. 

\begin{table}[htpb!]
\centering
\begin{tabular}{c|c|c}
 Parameter & Meaning & value\\
\midrule
 $\epsilon$ & Latency period   & 1 day  \\
 $\sigma$ & Infectious period & 14 days   \\
 $\omega$ & Immunity period & 140 days  \\
 $\lambda$ & Survival parameter & 0.9973  \\
 $\xi$ & Immunization parameter & see  panel below*  \\
 $\mu$ & Mortality rate & $3.68 \times 10^{-5}$  \\
 $\eta$ & threshold to start an infection focus & 0.001 \\
 $\beta_k$ & transmission parameter & see lowest panel**  \\
\midrule
\end{tabular}
\\ \footnotesize * Immunization rate $\xi$ is based on the proportion of the whole population vaccinated in a day according to \cite{THLvac}.Note that the number of people attaining full immunity is assumed to be about half of those vaccinated. \\
\bigskip
\begin{tabular}{c|c|c}
Starting day & Ending day & $\xi$* \\
\midrule
0 & 295 & 0 \\
295 & 344 & $2.7 \times 10^{-4}$\\
344 & 400 & $1.1 \times 10^{-3}$ \\
400 & 491 & $1.9 \times 10^{-3}$ \\
491 & 533 & $7.5 \times 10^{-4}$ \\
533 & 715 & $2.0 \times 10^{-4}$ \\
715 & end & $1.1 \times 10^{-5}$ \\
\midrule
\end{tabular}
\\ \footnotesize **  In the simulations the transmission parameter $\beta$ depends on the dominant strain ($k$) of the COVID-19 virus in Finland. Each significant strain has its period of dominance, which is based on the data from \cite{covariants}. See details below. \\
\bigskip
\begin{tabular}{c|c|c|c}
COVID Main Variant(s)    &  Starting day & Ending day & $\beta_{k}$* \\
\midrule
Initial  &  0 & 150 & 0.91 \\
EU1,EU2  &  150 & 330 & 1.1 \\
Alpha  &  330 & 520 & 2.3 \\
Delta &  520 & 670 & 4.5 \\
Omicron (21K)  &  670 & 720 & 10 \\
Omicron (21L)  &  720 & end & 7 \\
\midrule
\end{tabular}
\caption{\label{tab1} SEIRS model biological parameters. The two upper panels include epidemiological parameters  common to  all strains. The last panel shows the parameters that differ among strains.}
\label{table-biol}
\end{table} 

\newpage 

\section{The domain, setup and the general course of the simulations}

\label{simdom}

\begin{figure}[h]
\centering
\epsfxsize = 0.9\columnwidth \epsffile{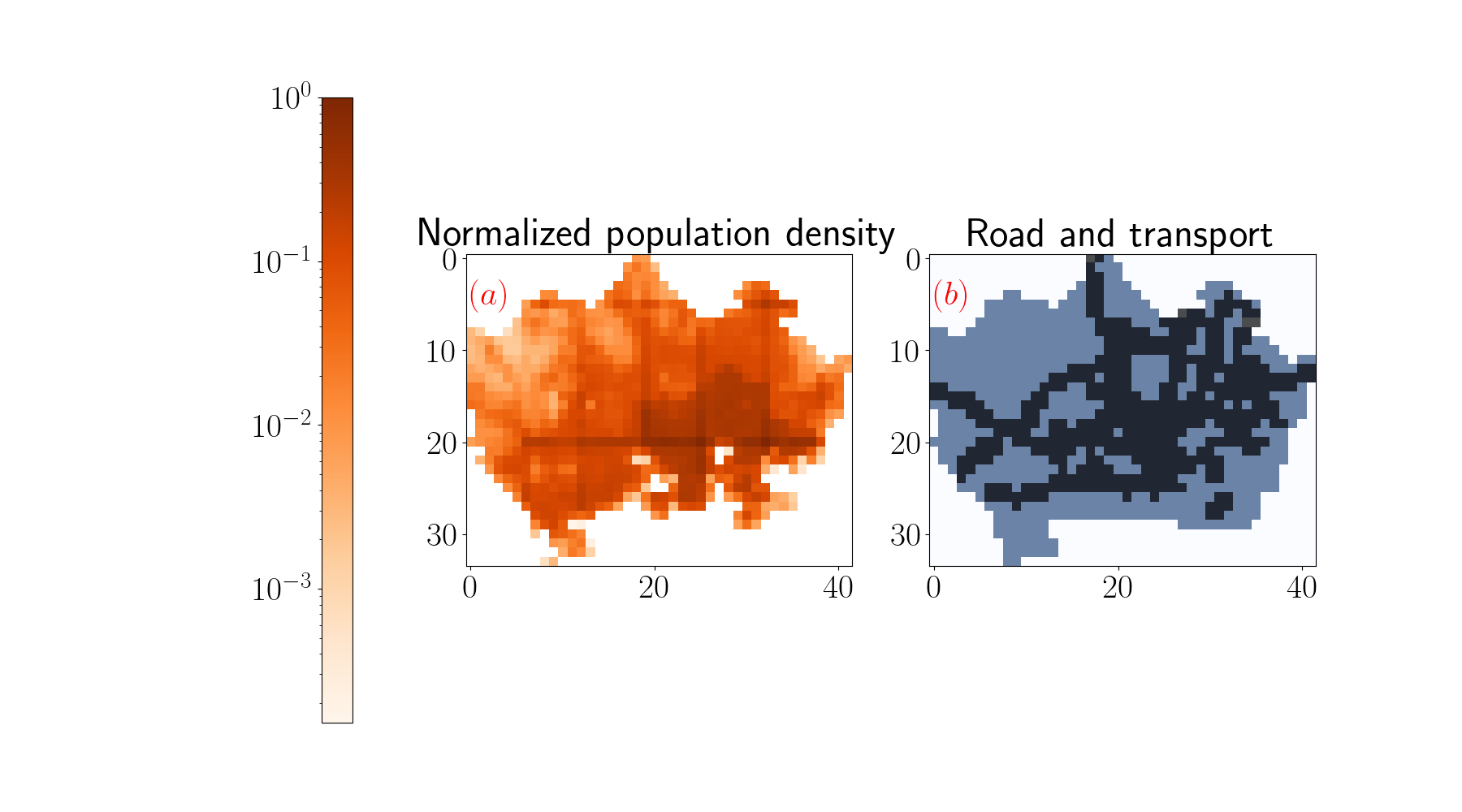}
\caption{The geographical domain used in the simulations. (a) The population density normalised between $0$ and $1$. (b) The public transport network in the simulations. }
\label{map}
\end{figure}

The geographical representation of the Helsinki metropolitan area as a square grid used in our simulations is visualised in Fig. \ref{map}, in which panel (a) shows the normalised population density \cite{worldpop:fin}, while panel (b) shows the public transport and road network based on the Google Maps digital map. The length scale of the square grid is $1$\mjk{km} $\times 1$\mjk{km}, which we chose as a compromise between two competing needs of the BTH--SEIRS-model: on one hand the grid needs enough cells so that the mobility of the populations can play a meaningful role in the simulations, which requires the length scale to be relatively small when compared with the full area. On the other hand, an individual square must cover an area large enough so that a person living in it would not ordinarily travel outside of it in his/her day to day business, excluding the commute to work. Without this size restriction, populations of different squares would effectively mix, which is something we want to avoid in the model. 

With our chosen length scale comes some challenges related to the generally small size of the capital city area and its relatively high population numbers. In total, the simulation domain consists of a square grid of $42$ times $34$ squares, or $1428$ squares, of which $877$ have non-zero population densities making up the population agents of our simulations. As the population density data in \cite{worldpop:fin} have a grid of approximately $1.5$\mjk{km} $\times 1.5$\mjk{km} in scale, it had to be resampled to fit the smaller one used in the simulations. Similarly, the road and transport network covers much of the simulation area, which is not that unrealistic considering that much of the capital city area is naturally urban with many mobility options both public and private: the metro, busses, trams, roads and streets for private cars or walking and cycling. 

The simulation is initialized at $t=0$, which corresponds to the 1st of March 2020, setting the infection rate at $\eta$ and the number of susceptible people at $1 - \eta$ in the geographical cell with the highest population density, along with a few other cells. The numbers of susceptible, exposed, infected and recovered people in each cell is updated at each simulation step by the SEIRS equations \ref{SEIRSeq1} and \ref{SEIRSeq2}. The epidemic may spread to new cells by any of the three forms of popular mobility presented in section \ref{SEIRScom} at each timestep, and once it does, its evolution proceeds in the same way as in the initially infected cells. The model parameters related to mobility are summarised in Table \ref{Mobparam}. From the total of four mobility parameters, two ($\gamma$,$KT$) govern the stochastic spreading through long distance trips, while the other two ($v_0$, $v_{max}$) couple the SEIRS and BTH parts of the model in accordance with Eq. \ref{vtdef}. Due to the small geographical scale of the present work and the much more dense road network in the Helsinki metropolitan area, the numerical values of the parameters $v_0$ and $v_{max}$ are much lower than those we used in our earlier study of Spain \cite{SpainApp}. As for the stochastic spreading,  geographical cells with population density below the threshold $\gamma$ can become infected due to stochastic long distance trips with probability $exp(-1/KT)$. 

The role of the value parameters is to modify the probability of epidemic spreading between cells via Eq. \ref{vtdef}, which defines the actual mobility $v^t_i$ in the cell inhabited by the population agent $i$. $v^t_i$ gives the probability of the epidemic spreading between neighbouring cells and those connected with the transport network, although in the latter case mobility is modified by the population densities of the connected cells. Different configurations of the value parameters can have very different results, as in an earlier study \cite{Somat} we have observed both fast and slow spreading patterns. $v^t_i$ also controls the local stochasticity mechanic, which turns out to be especially influential in this particular case: When fitting for the infection rates and economic activities, we find that in the general course of the simulations the epidemic will spread to all cells in relatively short amounts of time due to traffic connections and high population densities. Once all cells are infected, local stochasticity will be the only effect that modifies the infection rates in the cells. 

\begin{table}[htpb!]
\centering
\begin{tabular}{c|c|c}
 Parameter & Meaning & value\\
\midrule
 $KT$ & Long distance spreading parameter inspired by thermal noise & 0.22 \\
 $exp(-1/KT)$ & Probability of long range spreading & $\approx 0.01$ \\
 $\gamma$ & Maximum population density for long distance spreading & 0.05 \\
 $v_0$ & Normal mobility & 0.14 \\
 $v_{max}$ & Maximum mobility reduction & 0.13 \\
\midrule
\end{tabular}
\caption{Mobility parameters of the BTH--SEIRS model. }
\label{Mobparam}
\end{table}

\end{document}